\documentclass[prl,twocolumn,amsmath,amssymb,aps, superscriptaddress%, shortbibliography
]{revtex4-2}
\usepackage{amsmath} 
\usepackage{graphicx}
\usepackage{braket}
\usepackage{xcolor}
\usepackage[normalem]{ulem}
\usepackage{hyperref}
\usepackage{xcolor}
\usepackage{soul}
\begin{document}

\title{Interaction induced splitting of Dirac monopoles in the topological Thouless pumping of strongly interacting Bosons and SU($N$) Fermions}%
\author{Hei Lam} 
\affiliation{Department of Physics, The Chinese University of Hong Kong, Hong Kong SAR, China}
\author{Yangqian Yan}
\email{yqyan@cuhk.edu.hk}
\affiliation{Department of Physics, The Chinese University of Hong Kong, Hong Kong SAR, China}
\affiliation{The Chinese University of Hong Kong Shenzhen Research Institute, 518057 Shenzhen, China
}%

\begin{abstract}
Motivated by the observation of the breakdown of quantization for the Thouless pump in the presence of strong interaction by ETH [Walter et. al. Nat. Phys. 19, 1471 (2023), Viebahn et. al. arXiv:2308.03756], we study the interplay of strong interaction and topology in the (1+1)-dimensional interacting Rice-Mele model. We point out that the quantization of the interacting Thouless pump is dictated by the Chern number, i.e., the Dirac monopoles enclosed by the generalized Brillouin zone of the many-body wave function. By analyzing the change of location monopoles due to interaction, we predict the Thouless charge pump for strongly interacting Bose and SU($N$) Fermi gases in optical lattices and explain the ETH experiment. 

\end{abstract}

\maketitle

Ultracold atoms have been used to simulate non-interacting topological states of matter characterized by the first Chern number in optical lattices including Harper's model that exist in nature and Haldane model which is hard to attain~\cite{dasRealizingHarperModel2019,liuGeneralizedHaldaneModels2018,miyakeRealizingHarperHamiltonian2013}. Laughlin's state as a variational ansatz explains many states with fractional fillings. Among them, the two-body bosonic state has been successfully simulated in twisted optical cavities~\cite{clarkObservationLaughlinStates2020} and recently in ultracold atoms in optical lattices~\cite{taiMicroscopyInteractingHarper2017,leonardRealizationFractionalQuantum2023}. Yet, the many-body version remains to be simulated.

The adiabatic dynamics of the one-dimensional (1D) Rice-Mele model maps to a 2D Chern insulator via dimensional reduction~\cite{krausTopologicalEquivalenceFibonacci2012, qiTopologicalFieldTheory2008}. Thus, the dynamics are also governed by the Chern number, which is a closed surface integral of the Berry curvature over a generalized Brillouin zone spanned by quasi-momentum and time \cite{thoulessQuantizedHallConductance1982, xiaoBerryPhaseEffects2010,pekolaSingleelectronCurrentSources2013, QuantalPhaseFactors}. 

While the non-interacting quantum Thouless pump of the Rice-Mele model has been implemented in numerous quantum simulation platforms, including ultracold atoms~\cite{nakajimaTopologicalThoulessPumping2016, lohseThoulessQuantumPump2016, cooperTopologicalBandsUltracold2019}, nitrogen-vacancy centers in diamonds~\cite{maExperimentalObservationGeneralized2018}, and photonic systems~\cite{ozawaTopologicalPhotonics2019}, among others~\cite{citroThoulessPumpingTopology2023}, the interacting Thouless pumping, especially for bosons, has mostly been investigated theoretically~\cite{niuQuantisedAdiabaticCharge1984,bergQuantizedPumpingTopology2011,qianQuantumTransportBosonic2011,wangTopologicalChargePumping2013,zengFractionalChargePumping2016,nakagawaBreakdownTopologicalThouless2018, taoInteractioninducedTopologicalPumping2023}. In 2021 and 2022, the interaction-induced fractionalized Thouless pump for particular soliton mean-field wave function has been demonstrated in coupled waveguides and optical lattices experimentally~\cite{jurgensenQuantizedNonlinearThouless2021,fuNonlinearThoulessPumping2022}. 

At the mean-field level, it has been shown that the topological pumping is robust as long as the encircling is large to cover the superfluid phase~\cite{haywardTopologicalChargePumping2018,chongObservationNonequilibriumSteady2018}. In the strongly interacting and small encircling limit, the breakdown of Thouless pumping was predicted and observed in SU(2) Fermi gases~\cite{linInteractioninducedTopologicalBound2020,arguello-luengoStabilizationHubbardThoulessPumps2024}. The splitting of the topological defect was demonstrated recently~\cite{walterQuantizationItsBreakdown2023,viebahnInteractioninducedChargePumping2023,zhuReversalQuantizedHall2024}. Numerical simulations have been carried out~\cite{bertokSplittingTopologicalCharge2022}, but a comprehensive theoretical understanding of the phenomena, especially the influence of the topology under strong interaction beyond the mean-field effect, remains elusive~\cite{rachelInteractingTopologicalInsulators2018, wangClassificationInteractingElectronic2014, junemannExploringInteractingTopological2017,deleseleucObservationSymmetryprotectedTopological2019,luTheoryClassificationInteracting2012,freedmanClassTinvariantTopological2004}.

In 2023, the interaction-induced two-body bosonic Thouless pump was proposed and realized~\cite{keMultiparticleWannierStates2017}. The quantized pumping was determined to be connected to the polarization or the center-of-mass displacement caused by the many-body wave function. An exact diagonalization is required in their scheme. Thus, it is challenging to generalize it to many-body systems.

This paper theoretically investigates the pumping dynamics of the strongly interacting Rice-Mele model of bosons and SU($N$) fermions and provides theoretical explanations for the puzzles observed in recent experiments. Our key findings are as follows. (i) We obtained the exact locations of the Dirac monopoles of the $N$ bosons per unit cell as a function of interaction strength and $N$. These monopoles completely dictated the pumping in arbitrary encircling paths, which agrees with mean-field results in the literature in the weakly interacting limit. (ii) We extend the conclusions to SU($N$) fermions and point out that charge pumping is flavor-independent, which explains that quantized pumping is only observed in the first cycle. (iii) We restore pumping over many cycles for a particular flavor by introducing a staggered magnetic field. Our predictions are validated by time-evolved block decimation (TEBD), which explains recent ETH experiments and could be verified in both bosonic and SU($N$) fermionic systems.

\textit{Pumping in non-interacting systems.---}We consider the Rice-Mele model described by the time-dependent Hamiltonian,
\begin{equation}
\hat{H}_0(t) = \sum_{i} -J_{i}(t) (\hat{a}_{i}^{
\dag}\hat{a}_{i+1} + \text{h.c}) + \Delta_{i}(t)\hat{n}_{i} ,
\label{eq1}
\end{equation}
where $J_{i}(t)$ and $\Delta_{i}(t)$ are the hopping term and the on-site potential at the $i$-th site; generic adiabatic pumping paths are considered, but for simplicity, we focus our numerical verification on circular paths, e.g., $J_{i}(t) = J_0[1+(-1)^{i}\cos(\frac{2\pi}{T}t)]$ and $\Delta_i(t) = (-1)^{i}\Delta_{0}\sin(\frac{2\pi}{T}t)$. 

In the momentum space, we parameterize the Hamiltonian as 
\begin{equation}
    \hat{H}(k,t) = B_x(k,t) \sigma_x +B_y(k,t) \sigma_y +B_z(t) \sigma_z.
    \label{momentum}
\end{equation}
The total particle pumped in a cycle is determined by the Chern number, i.e., the integration of the Berry curvature defined within the generalized Brillouin zone, which implies an evenly occupied lowest band. For fermions, this is guaranteed by fermionic statistics with integer filling per unit cell. While for bosons, this could also be achieved by preparing an initial state using the maximally localized Wannier wave function, which is the equal-weight summation of every possible eigen state in the quasi-momentum space.

The generalized Brillouin is torus-like embeded in a 3D parameter space spanned by $B_x$, $B_y$, and $B_z$. Due to the inversion symmetry, to obtain the Chern number, we only have to evaluate the Berry curvature within half of the $\mathbb{T}^{2}$ torus. Thus, the Stocks theorem reduces the Chern number to the Zak phase difference of $H_0(k=0)$ and $H_0(k=\pi)$. Since the band degeneracy only occur at $k=\pi$ and not at $k=0$, 
the calculation in the 3D parameter space reduces to the winding number around the band touching point in the 2D parameter space spanned by $\delta= J_{1}-J_{2}$ and $\Delta$. This band touching point, which is located at the origin of the 2D parameter space for the non-interacting system, exactly corresponds to the location of the topological defect in the 3D parameter space.

\textit{Pumping model in interacting bosonic system.---}We consider onsite interactions,
$\sum_i\frac{U}{2}\hat{n}_{i}(\hat{n}_{i}-1)$.
In principle, the quantized Thouless pumping carries through to interacting systems as long as the initial wave function is evenly distributed over the generalized Brillouin zone defined by the total quasimomentum and the system does not pass through any gapless phases while slowly increasing the interaction strength.  When $t=0$, the intercell hopping is zero, and the band is completely flat.  Thus it is easy to construct the localized Wannier orbital with the uniform weight of momentum $k$. However, calculating the Berry curvature and the many-body wave function becomes challenging.  Instead of calculating the Berry curvature for the many-body wave functions, we calculate the topological defects' charge in the 3D parameter space and the number of times the generalized Brillouin zone wraps around the defects. 
Note that since the inversion symmetry is not broken, the Chern number reduces to the winding number in the 2D $\delta-\Delta$ parameter space. Thus, the calculation is reduced to searching for the band degeneracy points in the 2D parameter space and calculating their corresponding charge. Our idea could be thought of as the generalization of the band inversion theory to the interacting systems~\cite{zhangUnconventionalFloquetTopological2022}, and has been used in obtaining interacting displaced Yang monopoles in a 5D parameter space of synthetic dimensions~\cite{yanYangMonopolesEmergent2018}.

\begin{figure}
    \centering
    \includegraphics[scale=0.82]{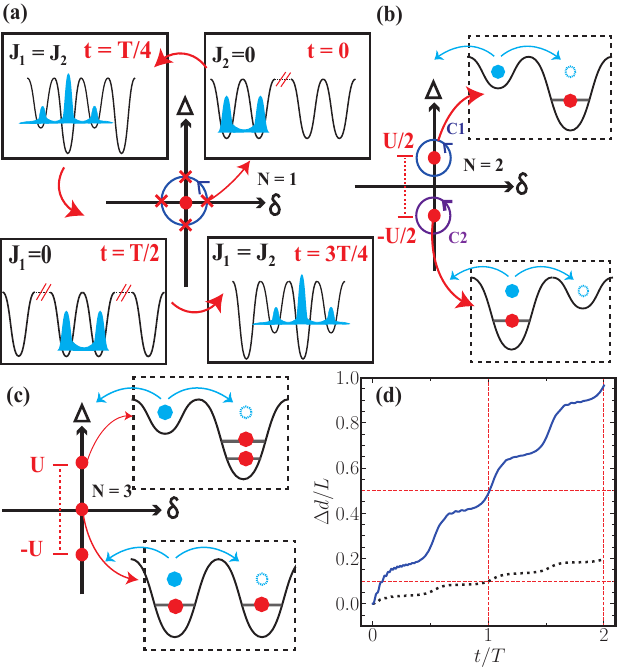}
    \caption{ (a-c) Schematic diagram of the location of topological defect points obtained from the effective Hamiltonian in the 2D parameter space under different particle filling in each unit cell. (d) The relative shift of the center of mass of bosons during an adiabatic 2-cycle pumping process. The blue solid line represents the case of half of the unit cells in the middle of the system, which are filled with two bosons, while the other is only filled with one. The black dotted line represents the case of a single boson pumping in the background of one particle filling per cell. The horizontal dashed lines mark the predictions from the quantized Thouless pump.}
    \label{fig:1}
\end{figure}

\textit{Effective single-particle Hamiltonian.---}For simplicity, we consider a single boson hopping in the presence of $N$ particles per unit cell. The ground state and energies could easily be written down on the $B_z$ axis. Since the states $\hat{a}^{\dag}_{i}\ket{\text{G.S}}$ and $\hat{b}^{\dag}_{i}\ket{\text{G.S}}$ becomes degenerate where $\ket{\text{G.S.}} = \prod_{i} ( \hat{a}^{\dag}_{i} )^{q}( \hat{b}_{i}^{\dag} )^{N-q} \ket{0} $ and $\ket{0}$ is the vacuum state, we find $N+1$ band touching points with the $q$-th point located at $(\delta, \Delta)=(0,\frac{U(2q-N)}{2})$, where $q=0,1,\cdots,N$.
To find the charge of these monopoles, we design an infinitely small trajectory that wraps around it. In this limit, we write down an effective Hamiltonian based on the degenerate perturbation theory using the aforementioned basis. We find that the new effective Hamiltonian is equivalent to particle hopping in the presence of $N$ particles fixed at the two sites in every unit cell except that the hopping amplitude is enhanced by the Bose enhancement factors $\sqrt{(N+1-q)(q+1)}$. The Chern number of the monopole turns out to be insensitive to the hopping amplitude as it only distorts the equi-Berry-curvature surface without changing the charge. We explicitly show the effective Hamiltonian and confirm our degenerate perturbation theory by comparing the spectrum using exact diagonalization in a small system (Supplementary Material~\cite{supnote}).

In order to generalize the non-interacting Thouless pump to many-body wave functions, we consider a system with uniform particle filling in each unit cell, leaving a single cell at the boundary where a particle has been removed. The Berry curvature of the many-body wave function is now defined on a Brillouin zone spanned by multi-particle quasi-momentum $K$ and time. Following a similar methodology, the effective Hamiltonian is derived by selecting two Bloch's states generated by the seed state in proximity to the monopole for perturbation, where the seed state denoted as $\{ \hat{A}^{\dagger}\ket{\text{G.S.}}, \hat{B}^{\dagger}\ket{\text{G.S.}} \}$, where $\hat{A}^{\dagger}$ and $\hat{B}^{\dagger}$ are creation operators defined by $\hat{A}^{\dagger} = \prod_{i=1}^{L-1}\hat{a}_{i}^{\dagger}$ and $\hat{B}^{\dagger} = \prod_{i=1}^{L-1}\hat{b}_{i}^{\dagger}$, respectively. The effective Hamiltonian exhibits a similar structure to the single-particle one, with distinctions in on-site energy and the Bose enhancement factor. However, it is the monopole, characterized by the effective Hamiltonian of the many-body system, that governs the group of $L - 1$ particles pumping, as opposed to the quantized pumping of a single particle. In the thermodynamic limit, where $L \rightarrow \infty$, the system approaches a uniformly filled system.

To conclude, we observe that $N$ copies of monopole with unit charge are evenly distributed along the $B_z$ or $\Delta$ axis and centered at the origin depending on the even-odd filling as shown in Fig.~\ref{fig:1}.
Since all the monopoles for the interacting systems smoothly connect to those for the non-interacting systems as the interaction strength decreases to zero, we believe all the monopoles have been located. The charge pumped in a period could then be easily calculated by counting the number of times the Brillouin wraps around the monopoles in 3D parameter space or simply the winding number around these defect points in the 2D parameter space.

Now, we consider the case with $m$ additional free bosons in the presence of $N$ particles per unit cell.

The generalized Brillouin zone may not be torus-like in the 3D parameter space. For example, for fully filled bands, the total quasi-momentum is single-valued, i.e., the torus-like structure in the 3D parameter space is squashed into a ring. Nevertheless, we find the winding number description is still robust.

\textit{Arbitrary adiabatic path.---}We use TEBD to verify our claims~\cite{fishmanITensorSoftwareLibrary2022}. We prepared $L/4$ particles on top of a $L / 2$ filled layer in the middle of the system and performed an imaginary-time TEBD (iTEBD) to obtain the system's ground state. To achieve the fully occupied band condition for the pumping, we prepare the initial state by starting the simulation from $t=0$ dimerized phase in which the inter-cell hopping is turned off during the iTEBD so that the center of mass (C.M.) of the system is highly localized and the density in the quasi-momentum will evenly be occupied in the beginning  (Supplementary Material~\cite{supnote}).
We evaluate the relative shift of C.M., $\Delta d=\text{C.M.(t)} - \text{C.M.(t=0)}$, where
$\text{C.M.}(t) = \langle\Psi(t)| \sum_{i}\frac{i}{L}\hat{n}_{i} | \Psi(t)\rangle.$
 We use the $T = 30/J$ for all the calculations to ensure the adiabaticity during the simulation. The time step size is chosen to be $0.005/J$. As shown in Fig.~\ref{fig:1}(d), we perform the calculation in trajectories along $(\delta,\Delta ) = (2\text{cos}(\frac{2\pi}{T}t) ,\Delta_{i}(t) + \frac{U}{2} ) $ (C1 in Fig.~\ref{fig:1}) centering the monopole calculated from the effective Hamiltonian as indicated in Fig.~\ref{fig:1}(b). A unit pumping was observed throughout the first pumping cycle. The charge pump is not sensitive to the initial condition in the 2D parameter space as long as we start from a maximally localized Wannier state. We mostly pick the dimerized state with the inter-cell hopping turned off, and the intra-cell hopping reaches its maximum value simply because the initial state is easy to prepare in simulations. We observe a quantized charge pump. Due to the finite size effect, the pumping is not quantized after many cycles. This is simply because of the reflection of a portion of the wave packet that already reached the edge of the box with open boundary condition. Adding a gradient field to the lattice~\cite{liuCorrelatedTopologicalPumping2023,wannierWaveFunctionsEffective1960,hartmannDynamicsBlochOscillations2004,holthausLocalizationEffectsAcdriven1996,dunlapDynamicLocalizationCharged1986},
$H_{f} = \sum_{i}\omega_{f} i \hat{n}_{i},$ fixes this numerical issue.
This technique has been for non-interacting system which leaves the topological charge pump unchanged but prevent the spread of the wave function.
For the strongly interacting systems, we also observe that the wave packet do not spread, and the quantized pumping is preserved after many cycles.

\textit{SU($N$) fermion pumping.---}We now generalize our theory to SU($N$) fermions and consider and Rice-Mele model with onsite interaction, 
\begin{equation}
\begin{split}
\hat{H} =& \sum_{i,\alpha } -J_{i}(t) (\hat{c}_{\alpha,i}^{\dag}\hat{c}_{\alpha,i+1} + \text{h.c}) + \\
&\sum_{i} \bigg(\Delta(t)\hat{n}_{i} + \sum_{\beta \neq \alpha } U\hat{n}_{\alpha, i}\hat{n}_{\beta,i}\bigg ).  
\end{split}
\end{equation}
As we will show below, the inclusion of this extra flavor index introduces a $N$-fold degeneracy in the limit $J_i\to 0$, which breaks the simple picture of Dirac monopoles.

For conciseness, we initiate our exposition with SU(2) and SU(3) models to elucidate the pumping mechanism in fermionic systems. The same analysis for bosons could be generalized to fermions. We first focus on two cases: (i) One spin-down particle per unit cell with an extra spin-up particle throughout the entire lattice. (ii) One spin-down and one spin-down particle per unit cell with one spin-up defect throughout the entire lattice. In both cases, the multi-particle Brillouin zone is well defined and the discussion for bosons could be generalized. Furthermore, in the thermodynamic limit, case (ii) approaches the fully filled scenario.

We start from the SU(2) fermions. Due to the added spin degree of freedom, the 2-fold degeneracy we observe at the band-touching point no longer holds. To demonstrate it, we employ exact diagonalization on a system comprising two unit cells, populated with two spin-up particles and one spin-down particle, under periodic boundary conditions. The system is projected into the quasi-momentum space in the limit where $J \ll U, \Delta$. Our findings reveal that the band crossing in the vicinity of the monopole is characterized by a 4-fold degeneracy. This could be understood in 2 folds: (i) the charge pump is spin insensitive; (ii) we have the freedom of coloring each particle with a particular spin for some instantaneous eigen states.
Upon dynamically evolving the system around this point of degeneracy, we note that at time $t = 0$, the ground band is unique. However, at $t = \frac{3T}{4}$, the system confronts an ambiguity in the evolutionary selection of states, as the ground-state band manifests a threefold degeneracy. The degenerate eigen state subspace is spanned by a linear combination of Bloch's basis, $\ket{k, r} = \frac{1}{\sqrt{L}}\sum_{p=1}^{2} e^{\mathbf{i}k2p}\hat{T}^{p}\ket{r}$, where $\ket{r}$ comprising $\{\ket{1} = \hat{c}_{\uparrow,1}^{\dagger}\hat{c}_{\uparrow,4}^{\dagger}\hat{c}_{\downarrow,2}^{\dagger}\ket{0}, \ket{2} = \hat{c}_{\uparrow,1}^{\dagger}\hat{c}_{\uparrow,2}^{\dagger}\hat{c}_{\downarrow,4}^{\dagger}\ket{0} , \ket{3}=\hat{c}_{\uparrow,2}^{\dagger}\hat{c}_{\uparrow,4}^{\dagger}\hat{c}_{\downarrow,1}^{\dagger}\ket{0} $). Consequently, the conventional perturbation methodology for the 2-band Hamiltonian ceases to be valid, and we do not expect a Dirac monopole and quantized pumping. This is substantiated by our TEBD simulation [dash-dotted line in Fig.~\ref{su2su3}(a)], where almost quantized pumping is observed in the first cycle but not the second cycle.

To restore the quantized pumping after the first cycle, we lift the triple degeneracy. This is achieved by introducing an additional staggered Zeeman field~\cite{bertokSplittingTopologicalCharge2022,citroThoulessPumpingTopology2023},
$
    H_{\text{Zeeman}} = B \sum_{i} (-1)^{i} \hat{S}^{z}_{s},
$
where $\hat{S}^{z}_{s}$ is the projection of spin-$s$ operator with $s=(N-1)/2$. This allows us to construct an effective Hamiltonian from a smaller subspace by selecting the lowest non-degenerate band state for the perturbation. Thus, the multi-band degeneracy is lifted and the effective Hamiltonian describing a Dirac monopole is recovered near the location of topological defects in the parameter space. Inspecting the state at $t=T/4$, we find that the wave function of the lowest two non-degenerate energy bands are contributed by Bloch's state generated by seed states $ \ket{1} = \hat{c}_{\uparrow,2}^{\dagger}\hat{c}_{\uparrow,4}^{\dagger}\hat{c}_{\downarrow,2}^{\dagger}\ket{0} $ and $\ket{2} = \hat{c}_{\uparrow,2}^{\dagger}\hat{c}_{\uparrow,4}^{\dagger}\hat{c}_{\downarrow,1}^{\dagger}\ket{0} $ independently. All spin-up particles are fixed in the lower sub-lattice sites and occupy the lowest energy state, while the spin-down particle occupies either lower or higher potential sites.

The dynamics could be understood as a spin-sensitive topological charge pump. This is because there exist two types of monopoles in the system according to their corresponding effective Hamiltonian, i.e., monopoles for spin-up or spin-down. As long as the initial state near the monopole is prepared in the lowest energy band state, the corresponding spin is pumped in a quantized way through the adiabatic cycle. For example, in Fig.~\ref{su2su3}(a), the top (bottom) defect corresponds to the pumping of spin-up (spin-down). Solid, dashed, and dotted lines in Fig.~\ref{su2su3}(a) shows the center of mass shift for the total, spin-up, and spin-downs as a function of time while winding around the top defect. We observe that the spin-up is pumped and the spin-down is unchanged.

Repeating the same calculation for the SU(3) fermions, in the presence of a staggered Zeeman field, gives rise to 3 monopoles located at $(\delta,\Delta)=( 0, U - B)$, $( 0, -U + B)$ and $(0,0)$. We find that each monopole corresponds to the pumping of a specific flavor. Generalizing to $SU(N)$ fermions, we find $N$ monopoles, similar to the structure of $N$ boson per unit cell.

\begin{figure}
    \centering
    \includegraphics[scale=0.86]{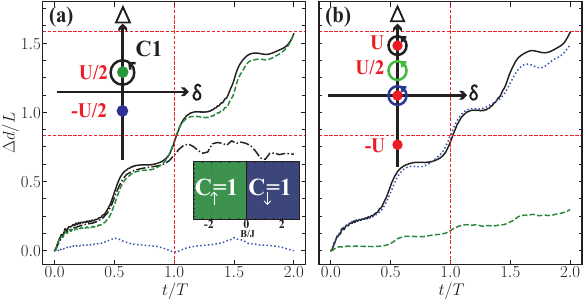}
    \caption{The relative shift of the center of mass of the SU(2) and SU(3) models with a defect filling in the end of the boundary along different trajectories in the parameter space. (a). The pumping of SU(2) system along the path C1. The black solid (dashed-dotted ) line represents the total relative shift with (without) the addition of the background staggered field while the green dashed (blue dotted) line corresponds to the relative shift of spin up (spin down) particle pumping in the background staggered field. (b). The pumping of the SU(3) model with the black solid line corresponding to the trajectory around the $\Delta=U$, and the green dashed line and blue dotted line correspond to the pumping of the trajectory around $\Delta = U/2 $ and $\Delta=0$.}
    \label{su2su3}
\end{figure}
To showcase the validity of our model, we conducted numerical simulations for both SU(2) and SU(3) systems. The simulations were performed on a system of size $L = 12$ for both SU(2) and SU(3) with the simulation period of $T = 30/J $. We traced a small circular trajectory $\text{C}_{1} $, as depicted in Fig.~\ref{su2su3}(a). We observe that, with the additional staggered Zeeman field, quantized pumping throughout the second cycle is restored, with only the spin-up particles being actively pumped [dashed lines in Fig.~\ref{su2su3}(a)]. We also observe an even-odd effect, same as that for bosonic systems, e.g. the defects for SU(3) are shifted by $U/2$ in the $\Delta$ axis [Fig.~\ref{su2su3}(b)]. Note that the charged pumped in the second cycle is smaller than the first cycle due to finite-size effect. This side effect could be eliminated by having less fermions on the top layer  (Supplementary Material~\cite{supnote}). We also note that the minor charge pump [dashed line Fig.~\ref{su2su3}(b)] while winding around $\Delta=U/2$ and $\delta=0$ for the SU(3) fermions is due to the quantized pumping of the hole, which disappears in the thermodynamic limit. Though only simulations encircle a single monopole is shown, we also perform simulations by varying the size of the circle  (Supplementary Material~\cite{supnote}). We verified that the charge pump is indeed equal to the total charge enclosed.
We expect the charge pump to be robust for a general SU($N$) fermi gas, but due to limited computational power, we did not perform simulators for a system with a larger number of flavors. 

In summary, we studied the strongly interacting Rice-Mele model. Instead of calculating the Berry curvature in the 2D generalized Brillouin zone, we opted to calculate the total charge enclosed. In the 3D parameter space, we located the topological defects and their charges, which completely determine the particles pumped in a cycle for strongly interacting Bosons and SU($N$) fermions. Our calculation goes beyond the mean-field level, explains the recent ETH experiment and could be verified in future ultracold atom experiments.

\begin{acknowledgments}
This work is supported by the Hong Kong RGC Early Career Scheme (Grants No. 24308323) and Collaborative Research Fund (Grant No. C4050-23GF), National Natural Science Foundation of China (Grant No. 12204395), and CUHK Direct Grant No. 4053676.
\end{acknowledgments}

\bibliographystyle{apstest}

\bibliography{reference,sup}

\begin{thebibliography}{52}%
\makeatletter
\providecommand \@ifxundefined [1]{%
 \@ifx{#1\undefined}
}%
\providecommand \@ifnum [1]{%
 \ifnum #1\expandafter \@firstoftwo
 \else \expandafter \@secondoftwo
 \fi
}%
\providecommand \@ifx [1]{%
 \ifx #1\expandafter \@firstoftwo
 \else \expandafter \@secondoftwo
 \fi
}%
\providecommand \natexlab [1]{#1}%
\providecommand \enquote  [1]{#1}%
\providecommand \bibnamefont  [1]{#1}%
\providecommand \bibfnamefont [1]{#1}%
\providecommand \citenamefont [1]{#1}%
\providecommand \href@noop [0]{\@secondoftwo}%
\providecommand \href [0]{\begingroup \@sanitize@url \@href}%
\providecommand \@href[1]{\@@startlink{#1}\@@href}%
\providecommand \@@href[1]{\endgroup#1\@@endlink}%
\providecommand \@sanitize@url [0]{\catcode `\\12\catcode `\$12\catcode `\&12\catcode `\#12\catcode `\^12\catcode `\_12\catcode `\%12\relax}%
\providecommand \@@startlink[1]{}%
\providecommand \@@endlink[0]{}%
\providecommand \url  [0]{\begingroup\@sanitize@url \@url }%
\providecommand \@url [1]{\endgroup\@href {#1}{\urlprefix }}%
\providecommand \urlprefix  [0]{URL }%
\providecommand \Eprint [0]{\href }%
\providecommand \doibase [0]{https://dx.doi.org}%
\providecommand \selectlanguage [0]{\@gobble}%
\providecommand \bibinfo  [0]{\@secondoftwo}%
\providecommand \bibfield  [0]{\@secondoftwo}%
\providecommand \translation [1]{[#1]}%
\providecommand \BibitemOpen [0]{}%
\providecommand \bibitemStop [0]{}%
\providecommand \bibitemNoStop [0]{.\EOS\space}%
\providecommand \EOS [0]{\spacefactor3000\relax}%
\providecommand \BibitemShut  [1]{\csname bibitem#1\endcsname}%
\let\auto@bib@innerbib\@empty
%</preamble>
\bibitem [{\citenamefont {Das}\ and\ \citenamefont {Christ}(2019)}]{dasRealizingHarperModel2019}%
  \BibitemOpen
  \bibfield  {author} {\bibinfo {author} {\bibfnamefont {K.~K.}\ \bibnamefont {Das}}\ and\ \bibinfo {author} {\bibfnamefont {J.}~\bibnamefont {Christ}},\ }\bibfield  {title} {\bibinfo {title} {Realizing the {{Harper}} model with ultracold atoms in a ring lattice},\ }\href {\doibase10.1103/PhysRevA.99.013604} {\bibfield  {journal} {\bibinfo  {journal} {Phys. Rev. A}\ }\textbf {\bibinfo {volume} {99}},\ \bibinfo {pages} {013604} (\bibinfo {year} {2019})}\BibitemShut {NoStop}%
\bibitem [{\citenamefont {Liu}\ \emph {et~al.}(2018)\citenamefont {Liu}, \citenamefont {Lin}, \citenamefont {Wang},\ and\ \citenamefont {Chen}}]{liuGeneralizedHaldaneModels2018}%
  \BibitemOpen
  \bibfield  {author} {\bibinfo {author} {\bibfnamefont {W.}~\bibnamefont {Liu}}, \bibinfo {author} {\bibfnamefont {Z.}~\bibnamefont {Lin}}, \bibinfo {author} {\bibfnamefont {Z.~D.}\ \bibnamefont {Wang}}, \ and\ \bibinfo {author} {\bibfnamefont {Y.}~\bibnamefont {Chen}},\ }\bibfield  {title} {\bibinfo {title} {Generalized {{Haldane}} models on laser-coupling optical lattices},\ }\href {\doibase10.1038/s41598-018-30503-9} {\bibfield  {journal} {\bibinfo  {journal} {Sci. Rep.}\ }\textbf {\bibinfo {volume} {8}},\ \bibinfo {pages} {12898} (\bibinfo {year} {2018})}\BibitemShut {NoStop}%
\bibitem [{\citenamefont {Miyake}\ \emph {et~al.}(2013)\citenamefont {Miyake}, \citenamefont {Siviloglou}, \citenamefont {Kennedy}, \citenamefont {Burton},\ and\ \citenamefont {{Wolfgang Ketterle}}}]{miyakeRealizingHarperHamiltonian2013}%
  \BibitemOpen
  \bibfield  {author} {\bibinfo {author} {\bibfnamefont {H.}~\bibnamefont {Miyake}}, \bibinfo {author} {\bibfnamefont {G.~A.}\ \bibnamefont {Siviloglou}}, \bibinfo {author} {\bibfnamefont {C.~J.}\ \bibnamefont {Kennedy}}, \bibinfo {author} {\bibfnamefont {W.~C.}\ \bibnamefont {Burton}}, \ and\ \bibinfo {author} {\bibnamefont {{Wolfgang Ketterle}}},\ }\bibfield  {title} {\bibinfo {title} {Realizing the {{Harper Hamiltonian}} with {{Laser-Assisted Tunneling}} in {{Optical Lattices}}},\ }\href {\doibase10.1103/PhysRevLett.111.185302} {\bibfield  {journal} {\bibinfo  {journal} {Phys. Rev. Lett.}\ }\textbf {\bibinfo {volume} {111}},\ \bibinfo {pages} {185302} (\bibinfo {year} {2013})}\BibitemShut {NoStop}%
\bibitem [{\citenamefont {Clark}\ \emph {et~al.}(2020)\citenamefont {Clark}, \citenamefont {Schine}, \citenamefont {Baum}, \citenamefont {Jia},\ and\ \citenamefont {Simon}}]{clarkObservationLaughlinStates2020}%
  \BibitemOpen
  \bibfield  {author} {\bibinfo {author} {\bibfnamefont {L.~W.}\ \bibnamefont {Clark}}, \bibinfo {author} {\bibfnamefont {N.}~\bibnamefont {Schine}}, \bibinfo {author} {\bibfnamefont {C.}~\bibnamefont {Baum}}, \bibinfo {author} {\bibfnamefont {N.}~\bibnamefont {Jia}}, \ and\ \bibinfo {author} {\bibfnamefont {J.}~\bibnamefont {Simon}},\ }\bibfield  {title} {\bibinfo {title} {Observation of {{Laughlin}} states made of light},\ }\href {\doibase10.1038/s41586-020-2318-5} {\bibfield  {journal} {\bibinfo  {journal} {Nature (London)}\ }\textbf {\bibinfo {volume} {582}},\ \bibinfo {pages} {41} (\bibinfo {year} {2020})}\BibitemShut {NoStop}%
\bibitem [{\citenamefont {Tai}\ \emph {et~al.}(2017)\citenamefont {Tai}, \citenamefont {Lukin}, \citenamefont {Rispoli}, \citenamefont {Schittko}, \citenamefont {Menke}, \citenamefont {{Dan Borgnia}}, \citenamefont {Preiss}, \citenamefont {Grusdt}, \citenamefont {Kaufman},\ and\ \citenamefont {Greiner}}]{taiMicroscopyInteractingHarper2017}%
  \BibitemOpen
  \bibfield  {author} {\bibinfo {author} {\bibfnamefont {M.~E.}\ \bibnamefont {Tai}}, \bibinfo {author} {\bibfnamefont {A.}~\bibnamefont {Lukin}}, \bibinfo {author} {\bibfnamefont {M.}~\bibnamefont {Rispoli}}, \bibinfo {author} {\bibfnamefont {R.}~\bibnamefont {Schittko}}, \bibinfo {author} {\bibfnamefont {T.}~\bibnamefont {Menke}}, \bibinfo {author} {\bibnamefont {{Dan Borgnia}}}, \bibinfo {author} {\bibfnamefont {P.~M.}\ \bibnamefont {Preiss}}, \bibinfo {author} {\bibfnamefont {F.}~\bibnamefont {Grusdt}}, \bibinfo {author} {\bibfnamefont {A.~M.}\ \bibnamefont {Kaufman}}, \ and\ \bibinfo {author} {\bibfnamefont {M.}~\bibnamefont {Greiner}},\ }\bibfield  {title} {\bibinfo {title} {Microscopy of the interacting {{Harper}}–{{Hofstadter}} model in the two-body limit},\ }\href {\doibase10.1038/nature22811} {\bibfield  {journal} {\bibinfo  {journal} {Nature (London)}\ }\textbf {\bibinfo {volume} {546}},\ \bibinfo {pages} {519} (\bibinfo {year} {2017})}\BibitemShut {NoStop}%
\bibitem [{\citenamefont {Léonard}\ \emph {et~al.}(2023)\citenamefont {Léonard}, \citenamefont {Kim}, \citenamefont {Kwan}, \citenamefont {Segura}, \citenamefont {Grusdt}, \citenamefont {Repellin}, \citenamefont {Goldman},\ and\ \citenamefont {Greiner}}]{leonardRealizationFractionalQuantum2023}%
  \BibitemOpen
  \bibfield  {author} {\bibinfo {author} {\bibfnamefont {J.}~\bibnamefont {Léonard}}, \bibinfo {author} {\bibfnamefont {S.}~\bibnamefont {Kim}}, \bibinfo {author} {\bibfnamefont {J.}~\bibnamefont {Kwan}}, \bibinfo {author} {\bibfnamefont {P.}~\bibnamefont {Segura}}, \bibinfo {author} {\bibfnamefont {F.}~\bibnamefont {Grusdt}}, \bibinfo {author} {\bibfnamefont {C.}~\bibnamefont {Repellin}}, \bibinfo {author} {\bibfnamefont {N.}~\bibnamefont {Goldman}}, \ and\ \bibinfo {author} {\bibfnamefont {M.}~\bibnamefont {Greiner}},\ }\bibfield  {title} {\bibinfo {title} {Realization of a fractional quantum {{Hall}} state with ultracold atoms},\ }\href {\doibase10.1038/s41586-023-06122-4} {\bibfield  {journal} {\bibinfo  {journal} {Nature}\ }\textbf {\bibinfo {volume} {619}},\ \bibinfo {pages} {495} (\bibinfo {year} {2023})}\BibitemShut {NoStop}%
\bibitem [{\citenamefont {Kraus}\ and\ \citenamefont {Zilberberg}(2012)}]{krausTopologicalEquivalenceFibonacci2012}%
  \BibitemOpen
  \bibfield  {author} {\bibinfo {author} {\bibfnamefont {Y.~E.}\ \bibnamefont {Kraus}}\ and\ \bibinfo {author} {\bibfnamefont {O.}~\bibnamefont {Zilberberg}},\ }\bibfield  {title} {\bibinfo {title} {Topological {{Equivalence}} between the {{Fibonacci Quasicrystal}} and the {{Harper Model}}},\ }\href {\doibase10.1103/PhysRevLett.109.116404} {\bibfield  {journal} {\bibinfo  {journal} {Phys. Rev. Lett.}\ }\textbf {\bibinfo {volume} {109}},\ \bibinfo {pages} {116404} (\bibinfo {year} {2012})}\BibitemShut {NoStop}%
\bibitem [{\citenamefont {Qi}\ \emph {et~al.}(2008)\citenamefont {Qi}, \citenamefont {Hughes},\ and\ \citenamefont {Zhang}}]{qiTopologicalFieldTheory2008}%
  \BibitemOpen
  \bibfield  {author} {\bibinfo {author} {\bibfnamefont {X.-L.}\ \bibnamefont {Qi}}, \bibinfo {author} {\bibfnamefont {T.~L.}\ \bibnamefont {Hughes}}, \ and\ \bibinfo {author} {\bibfnamefont {S.-C.}\ \bibnamefont {Zhang}},\ }\bibfield  {title} {\bibinfo {title} {Topological field theory of time-reversal invariant insulators},\ }\href {\doibase10.1103/PhysRevB.78.195424} {\bibfield  {journal} {\bibinfo  {journal} {Phys. Rev. B}\ }\textbf {\bibinfo {volume} {78}},\ \bibinfo {pages} {195424} (\bibinfo {year} {2008})}\BibitemShut {NoStop}%
\bibitem [{\citenamefont {Thouless}\ \emph {et~al.}(1982)\citenamefont {Thouless}, \citenamefont {Kohmoto}, \citenamefont {Nightingale},\ and\ \citenamefont {{den Nijs}}}]{thoulessQuantizedHallConductance1982}%
  \BibitemOpen
  \bibfield  {author} {\bibinfo {author} {\bibfnamefont {D.~J.}\ \bibnamefont {Thouless}}, \bibinfo {author} {\bibfnamefont {M.}~\bibnamefont {Kohmoto}}, \bibinfo {author} {\bibfnamefont {M.~P.}\ \bibnamefont {Nightingale}}, \ and\ \bibinfo {author} {\bibfnamefont {M.}~\bibnamefont {{den Nijs}}},\ }\bibfield  {title} {\bibinfo {title} {Quantized {{Hall Conductance}} in a {{Two-Dimensional Periodic Potential}}},\ }\href {\doibase10.1103/PhysRevLett.49.405} {\bibfield  {journal} {\bibinfo  {journal} {Phys. Rev. Lett.}\ }\textbf {\bibinfo {volume} {49}},\ \bibinfo {pages} {405} (\bibinfo {year} {1982})}\BibitemShut {NoStop}%
\bibitem [{\citenamefont {Xiao}\ \emph {et~al.}(2010)\citenamefont {Xiao}, \citenamefont {Chang},\ and\ \citenamefont {Niu}}]{xiaoBerryPhaseEffects2010}%
  \BibitemOpen
  \bibfield  {author} {\bibinfo {author} {\bibfnamefont {D.}~\bibnamefont {Xiao}}, \bibinfo {author} {\bibfnamefont {M.-C.}\ \bibnamefont {Chang}}, \ and\ \bibinfo {author} {\bibfnamefont {Q.}~\bibnamefont {Niu}},\ }\bibfield  {title} {\bibinfo {title} {Berry phase effects on electronic properties},\ }\href {\doibase10.1103/RevModPhys.82.1959} {\bibfield  {journal} {\bibinfo  {journal} {Rev. Mod. Phys.}\ }\textbf {\bibinfo {volume} {82}},\ \bibinfo {pages} {1959} (\bibinfo {year} {2010})}\BibitemShut {NoStop}%
\bibitem [{\citenamefont {Pekola}\ \emph {et~al.}(2013)\citenamefont {Pekola}, \citenamefont {Saira}, \citenamefont {Maisi}, \citenamefont {Kemppinen}, \citenamefont {Möttönen}, \citenamefont {Pashkin},\ and\ \citenamefont {Averin}}]{pekolaSingleelectronCurrentSources2013}%
  \BibitemOpen
  \bibfield  {author} {\bibinfo {author} {\bibfnamefont {J.~P.}\ \bibnamefont {Pekola}}, \bibinfo {author} {\bibfnamefont {O.-P.}\ \bibnamefont {Saira}}, \bibinfo {author} {\bibfnamefont {V.~F.}\ \bibnamefont {Maisi}}, \bibinfo {author} {\bibfnamefont {A.}~\bibnamefont {Kemppinen}}, \bibinfo {author} {\bibfnamefont {M.}~\bibnamefont {Möttönen}}, \bibinfo {author} {\bibfnamefont {Y.~A.}\ \bibnamefont {Pashkin}}, \ and\ \bibinfo {author} {\bibfnamefont {D.~V.}\ \bibnamefont {Averin}},\ }\bibfield  {title} {\bibinfo {title} {Single-electron current sources: {{Toward}} a refined definition of the ampere},\ }\href {\doibase10.1103/RevModPhys.85.1421} {\bibfield  {journal} {\bibinfo  {journal} {Rev. Mod. Phys.}\ }\textbf {\bibinfo {volume} {85}},\ \bibinfo {pages} {1421} (\bibinfo {year} {2013})}\BibitemShut {NoStop}%
\bibitem [{\citenamefont {Berry}(1984)}]{QuantalPhaseFactors}%
  \BibitemOpen
  \bibfield  {author} {\bibinfo {author} {\bibfnamefont {M.~V.}\ \bibnamefont {Berry}},\ }\bibfield  {title} {\bibinfo {title} {Quantal phase factors accompanying adiabatic changes},\ }\href {\doibase10.1098/rspa.1984.0023} {\bibfield  {journal} {\bibinfo  {journal} {Proc. R. Soc. Lond.}\ }\textbf {\bibinfo {volume} {392}},\ \bibinfo {pages} {45} (\bibinfo {year} {1984})}\BibitemShut {NoStop}%
\bibitem [{\citenamefont {Nakajima}\ \emph {et~al.}(2016)\citenamefont {Nakajima}, \citenamefont {Tomita}, \citenamefont {Taie}, \citenamefont {Ichinose}, \citenamefont {Ozawa}, \citenamefont {Wang}, \citenamefont {Troyer},\ and\ \citenamefont {Takahashi}}]{nakajimaTopologicalThoulessPumping2016}%
  \BibitemOpen
  \bibfield  {author} {\bibinfo {author} {\bibfnamefont {S.}~\bibnamefont {Nakajima}}, \bibinfo {author} {\bibfnamefont {T.}~\bibnamefont {Tomita}}, \bibinfo {author} {\bibfnamefont {S.}~\bibnamefont {Taie}}, \bibinfo {author} {\bibfnamefont {T.}~\bibnamefont {Ichinose}}, \bibinfo {author} {\bibfnamefont {H.}~\bibnamefont {Ozawa}}, \bibinfo {author} {\bibfnamefont {L.}~\bibnamefont {Wang}}, \bibinfo {author} {\bibfnamefont {M.}~\bibnamefont {Troyer}}, \ and\ \bibinfo {author} {\bibfnamefont {Y.}~\bibnamefont {Takahashi}},\ }\bibfield  {title} {\bibinfo {title} {Topological {{Thouless}} pumping of ultracold fermions},\ }\href {\doibase10.1038/nphys3622} {\bibfield  {journal} {\bibinfo  {journal} {Nat. Phys.}\ }\textbf {\bibinfo {volume} {12}},\ \bibinfo {pages} {296} (\bibinfo {year} {2016})}\BibitemShut {NoStop}%
\bibitem [{\citenamefont {Lohse}\ \emph {et~al.}(2016)\citenamefont {Lohse}, \citenamefont {Schweizer}, \citenamefont {Zilberberg}, \citenamefont {Aidelsburger},\ and\ \citenamefont {Bloch}}]{lohseThoulessQuantumPump2016}%
  \BibitemOpen
  \bibfield  {author} {\bibinfo {author} {\bibfnamefont {M.}~\bibnamefont {Lohse}}, \bibinfo {author} {\bibfnamefont {C.}~\bibnamefont {Schweizer}}, \bibinfo {author} {\bibfnamefont {O.}~\bibnamefont {Zilberberg}}, \bibinfo {author} {\bibfnamefont {M.}~\bibnamefont {Aidelsburger}}, \ and\ \bibinfo {author} {\bibfnamefont {I.}~\bibnamefont {Bloch}},\ }\bibfield  {title} {\bibinfo {title} {A {{Thouless}} quantum pump with ultracold bosonic atoms in an optical superlattice},\ }\href {\doibase10.1038/nphys3584} {\bibfield  {journal} {\bibinfo  {journal} {Nat. Phys.}\ }\textbf {\bibinfo {volume} {12}},\ \bibinfo {pages} {350} (\bibinfo {year} {2016})}\BibitemShut {NoStop}%
\bibitem [{\citenamefont {Cooper}\ \emph {et~al.}(2019)\citenamefont {Cooper}, \citenamefont {Dalibard},\ and\ \citenamefont {Spielman}}]{cooperTopologicalBandsUltracold2019}%
  \BibitemOpen
  \bibfield  {author} {\bibinfo {author} {\bibfnamefont {N.~R.}\ \bibnamefont {Cooper}}, \bibinfo {author} {\bibfnamefont {J.}~\bibnamefont {Dalibard}}, \ and\ \bibinfo {author} {\bibfnamefont {I.~B.}\ \bibnamefont {Spielman}},\ }\bibfield  {title} {\bibinfo {title} {Topological bands for ultracold atoms},\ }\href {\doibase10.1103/RevModPhys.91.015005} {\bibfield  {journal} {\bibinfo  {journal} {Rev. Mod. Phys.}\ }\textbf {\bibinfo {volume} {91}},\ \bibinfo {pages} {015005} (\bibinfo {year} {2019})}\BibitemShut {NoStop}%
\bibitem [{\citenamefont {Ma}\ \emph {et~al.}(2018)\citenamefont {Ma}, \citenamefont {Zhou}, \citenamefont {Zhang}, \citenamefont {Li}, \citenamefont {Cheng}, \citenamefont {Geng}, \citenamefont {Rong}, \citenamefont {Shi}, \citenamefont {Gong},\ and\ \citenamefont {Du}}]{maExperimentalObservationGeneralized2018}%
  \BibitemOpen
  \bibfield  {author} {\bibinfo {author} {\bibfnamefont {W.}~\bibnamefont {Ma}}, \bibinfo {author} {\bibfnamefont {L.}~\bibnamefont {Zhou}}, \bibinfo {author} {\bibfnamefont {Q.}~\bibnamefont {Zhang}}, \bibinfo {author} {\bibfnamefont {M.}~\bibnamefont {Li}}, \bibinfo {author} {\bibfnamefont {C.}~\bibnamefont {Cheng}}, \bibinfo {author} {\bibfnamefont {J.}~\bibnamefont {Geng}}, \bibinfo {author} {\bibfnamefont {X.}~\bibnamefont {Rong}}, \bibinfo {author} {\bibfnamefont {F.}~\bibnamefont {Shi}}, \bibinfo {author} {\bibfnamefont {J.}~\bibnamefont {Gong}}, \ and\ \bibinfo {author} {\bibfnamefont {J.}~\bibnamefont {Du}},\ }\bibfield  {title} {\bibinfo {title} {Experimental {{Observation}} of a {{Generalized Thouless Pump}} with a {{Single Spin}}},\ }\href {\doibase10.1103/PhysRevLett.120.120501} {\bibfield  {journal} {\bibinfo  {journal} {Phys. Rev. Lett.}\ }\textbf {\bibinfo {volume} {120}},\ \bibinfo {pages} {120501} (\bibinfo {year} {2018})}\BibitemShut {NoStop}%
\bibitem [{\citenamefont {Ozawa}\ \emph {et~al.}(2019)\citenamefont {Ozawa}, \citenamefont {Price}, \citenamefont {Amo}, \citenamefont {Goldman}, \citenamefont {Hafezi}, \citenamefont {Lu}, \citenamefont {Rechtsman}, \citenamefont {Schuster}, \citenamefont {Simon}, \citenamefont {Zilberberg},\ and\ \citenamefont {Carusotto}}]{ozawaTopologicalPhotonics2019}%
  \BibitemOpen
  \bibfield  {author} {\bibinfo {author} {\bibfnamefont {T.}~\bibnamefont {Ozawa}}, \bibinfo {author} {\bibfnamefont {H.~M.}\ \bibnamefont {Price}}, \bibinfo {author} {\bibfnamefont {A.}~\bibnamefont {Amo}}, \bibinfo {author} {\bibfnamefont {N.}~\bibnamefont {Goldman}}, \bibinfo {author} {\bibfnamefont {M.}~\bibnamefont {Hafezi}}, \bibinfo {author} {\bibfnamefont {L.}~\bibnamefont {Lu}}, \bibinfo {author} {\bibfnamefont {M.~C.}\ \bibnamefont {Rechtsman}}, \bibinfo {author} {\bibfnamefont {D.}~\bibnamefont {Schuster}}, \bibinfo {author} {\bibfnamefont {J.}~\bibnamefont {Simon}}, \bibinfo {author} {\bibfnamefont {O.}~\bibnamefont {Zilberberg}}, \ and\ \bibinfo {author} {\bibfnamefont {I.}~\bibnamefont {Carusotto}},\ }\bibfield  {title} {\bibinfo {title} {Topological photonics},\ }\href {\doibase10.1103/RevModPhys.91.015006} {\bibfield  {journal} {\bibinfo  {journal} {Rev. Mod. Phys.}\ }\textbf {\bibinfo {volume} {91}},\ \bibinfo {pages} {015006} (\bibinfo {year} {2019})}\BibitemShut {NoStop}%
\bibitem [{\citenamefont {Citro}\ and\ \citenamefont {Aidelsburger}(2023)}]{citroThoulessPumpingTopology2023}%
  \BibitemOpen
  \bibfield  {author} {\bibinfo {author} {\bibfnamefont {R.}~\bibnamefont {Citro}}\ and\ \bibinfo {author} {\bibfnamefont {M.}~\bibnamefont {Aidelsburger}},\ }\bibfield  {title} {\bibinfo {title} {Thouless pumping and topology},\ }\href {\doibase10.1038/s42254-022-00545-0} {\bibfield  {journal} {\bibinfo  {journal} {Nat. Rev. Phys.}\ }\textbf {\bibinfo {volume} {5}},\ \bibinfo {pages} {87} (\bibinfo {year} {2023})}\BibitemShut {NoStop}%
\bibitem [{\citenamefont {Niu}\ and\ \citenamefont {Thouless}(1984)}]{niuQuantisedAdiabaticCharge1984}%
  \BibitemOpen
  \bibfield  {author} {\bibinfo {author} {\bibfnamefont {Q.}~\bibnamefont {Niu}}\ and\ \bibinfo {author} {\bibfnamefont {D.~J.}\ \bibnamefont {Thouless}},\ }\bibfield  {title} {\bibinfo {title} {Quantised adiabatic charge transport in the presence of substrate disorder and many-body interaction},\ }\href {\doibase10.1088/0305-4470/17/12/016} {\bibfield  {journal} {\bibinfo  {journal} {J. Phys. A}\ }\textbf {\bibinfo {volume} {17}},\ \bibinfo {pages} {2453} (\bibinfo {year} {1984})}\BibitemShut {NoStop}%
\bibitem [{\citenamefont {Berg}\ \emph {et~al.}(2011)\citenamefont {Berg}, \citenamefont {Levin},\ and\ \citenamefont {Altman}}]{bergQuantizedPumpingTopology2011}%
  \BibitemOpen
  \bibfield  {author} {\bibinfo {author} {\bibfnamefont {E.}~\bibnamefont {Berg}}, \bibinfo {author} {\bibfnamefont {M.}~\bibnamefont {Levin}}, \ and\ \bibinfo {author} {\bibfnamefont {E.}~\bibnamefont {Altman}},\ }\bibfield  {title} {\bibinfo {title} {Quantized {{Pumping}} and {{Topology}} of the {{Phase Diagram}} for a {{System}} of {{Interacting Bosons}}},\ }\href {\doibase10.1103/PhysRevLett.106.110405} {\bibfield  {journal} {\bibinfo  {journal} {Phys. Rev. Lett.}\ }\textbf {\bibinfo {volume} {106}},\ \bibinfo {pages} {110405} (\bibinfo {year} {2011})}\BibitemShut {NoStop}%
\bibitem [{\citenamefont {Qian}\ \emph {et~al.}(2011)\citenamefont {Qian}, \citenamefont {Gong},\ and\ \citenamefont {Zhang}}]{qianQuantumTransportBosonic2011}%
  \BibitemOpen
  \bibfield  {author} {\bibinfo {author} {\bibfnamefont {Y.}~\bibnamefont {Qian}}, \bibinfo {author} {\bibfnamefont {M.}~\bibnamefont {Gong}}, \ and\ \bibinfo {author} {\bibfnamefont {C.}~\bibnamefont {Zhang}},\ }\bibfield  {title} {\bibinfo {title} {Quantum transport of bosonic cold atoms in double-well optical lattices},\ }\href {\doibase10.1103/PhysRevA.84.013608} {\bibfield  {journal} {\bibinfo  {journal} {Phys. Rev. A}\ }\textbf {\bibinfo {volume} {84}},\ \bibinfo {pages} {013608} (\bibinfo {year} {2011})}\BibitemShut {NoStop}%
\bibitem [{\citenamefont {Wang}\ \emph {et~al.}(2013)\citenamefont {Wang}, \citenamefont {Troyer},\ and\ \citenamefont {Dai}}]{wangTopologicalChargePumping2013}%
  \BibitemOpen
  \bibfield  {author} {\bibinfo {author} {\bibfnamefont {L.}~\bibnamefont {Wang}}, \bibinfo {author} {\bibfnamefont {M.}~\bibnamefont {Troyer}}, \ and\ \bibinfo {author} {\bibfnamefont {X.}~\bibnamefont {Dai}},\ }\bibfield  {title} {\bibinfo {title} {Topological {{Charge Pumping}} in a {{One-Dimensional Optical Lattice}}},\ }\href {\doibase10.1103/PhysRevLett.111.026802} {\bibfield  {journal} {\bibinfo  {journal} {Phys. Rev. Lett.}\ }\textbf {\bibinfo {volume} {111}},\ \bibinfo {pages} {026802} (\bibinfo {year} {2013})}\BibitemShut {NoStop}%
\bibitem [{\citenamefont {Zeng}\ \emph {et~al.}(2016)\citenamefont {Zeng}, \citenamefont {Zhu},\ and\ \citenamefont {Sheng}}]{zengFractionalChargePumping2016}%
  \BibitemOpen
  \bibfield  {author} {\bibinfo {author} {\bibfnamefont {T.-S.}\ \bibnamefont {Zeng}}, \bibinfo {author} {\bibfnamefont {W.}~\bibnamefont {Zhu}}, \ and\ \bibinfo {author} {\bibfnamefont {D.~N.}\ \bibnamefont {Sheng}},\ }\bibfield  {title} {\bibinfo {title} {Fractional charge pumping of interacting bosons in one-dimensional superlattice},\ }\href {\doibase10.1103/PhysRevB.94.235139} {\bibfield  {journal} {\bibinfo  {journal} {Phys. Rev. B}\ }\textbf {\bibinfo {volume} {94}},\ \bibinfo {pages} {235139} (\bibinfo {year} {2016})}\BibitemShut {NoStop}%
\bibitem [{\citenamefont {Nakagawa}\ \emph {et~al.}(2018)\citenamefont {Nakagawa}, \citenamefont {Yoshida}, \citenamefont {Peters},\ and\ \citenamefont {Kawakami}}]{nakagawaBreakdownTopologicalThouless2018}%
  \BibitemOpen
  \bibfield  {author} {\bibinfo {author} {\bibfnamefont {M.}~\bibnamefont {Nakagawa}}, \bibinfo {author} {\bibfnamefont {T.}~\bibnamefont {Yoshida}}, \bibinfo {author} {\bibfnamefont {R.}~\bibnamefont {Peters}}, \ and\ \bibinfo {author} {\bibfnamefont {N.}~\bibnamefont {Kawakami}},\ }\bibfield  {title} {\bibinfo {title} {Breakdown of topological {{Thouless}} pumping in the strongly interacting regime},\ }\href {\doibase10.1103/PhysRevB.98.115147} {\bibfield  {journal} {\bibinfo  {journal} {Phys. Rev. B}\ }\textbf {\bibinfo {volume} {98}},\ \bibinfo {pages} {115147} (\bibinfo {year} {2018})}\BibitemShut {NoStop}%
\bibitem [{\citenamefont {Tao}\ \emph {et~al.}()\citenamefont {Tao}, \citenamefont {Huang}, \citenamefont {Niu}, \citenamefont {Zhang}, \citenamefont {Ke}, \citenamefont {Gu}, \citenamefont {Lin}, \citenamefont {Qiu}, \citenamefont {Sun}, \citenamefont {Yang}, \citenamefont {Zhang}, \citenamefont {Zhang}, \citenamefont {Zhao}, \citenamefont {Zhou}, \citenamefont {Deng}, \citenamefont {Hu}, \citenamefont {Hu}, \citenamefont {Li}, \citenamefont {Liu}, \citenamefont {Tan}, \citenamefont {Xu}, \citenamefont {Yan}, \citenamefont {Chen}, \citenamefont {Lee}, \citenamefont {Zhong}, \citenamefont {Liu},\ and\ \citenamefont {Yu}}]{taoInteractioninducedTopologicalPumping2023}%
  \BibitemOpen
  \bibfield  {author} {\bibinfo {author} {\bibfnamefont {Z.}~\bibnamefont {Tao}}, \bibinfo {author} {\bibfnamefont {W.}~\bibnamefont {Huang}}, \bibinfo {author} {\bibfnamefont {J.}~\bibnamefont {Niu}}, \bibinfo {author} {\bibfnamefont {L.}~\bibnamefont {Zhang}}, \bibinfo {author} {\bibfnamefont {Y.}~\bibnamefont {Ke}}, \bibinfo {author} {\bibfnamefont {X.}~\bibnamefont {Gu}}, \bibinfo {author} {\bibfnamefont {L.}~\bibnamefont {Lin}}, \bibinfo {author} {\bibfnamefont {J.}~\bibnamefont {Qiu}}, \bibinfo {author} {\bibfnamefont {X.}~\bibnamefont {Sun}}, \bibinfo {author} {\bibfnamefont {X.}~\bibnamefont {Yang}}, \bibinfo {author} {\bibfnamefont {J.}~\bibnamefont {Zhang}}, \bibinfo {author} {\bibfnamefont {J.}~\bibnamefont {Zhang}}, \bibinfo {author} {\bibfnamefont {S.}~\bibnamefont {Zhao}}, \bibinfo {author} {\bibfnamefont {Y.}~\bibnamefont {Zhou}}, \bibinfo {author} {\bibfnamefont {X.}~\bibnamefont {Deng}}, \bibinfo {author} {\bibfnamefont {C.}~\bibnamefont {Hu}}, \bibinfo {author} {\bibfnamefont {L.}~\bibnamefont {Hu}}, \bibinfo {author} {\bibfnamefont {J.}~\bibnamefont {Li}}, \bibinfo {author} {\bibfnamefont {Y.}~\bibnamefont {Liu}}, \bibinfo {author} {\bibfnamefont {D.}~\bibnamefont {Tan}}, \bibinfo {author} {\bibfnamefont {Y.}~\bibnamefont {Xu}}, \bibinfo {author} {\bibfnamefont {T.}~\bibnamefont {Yan}}, \bibinfo {author} {\bibfnamefont {Y.}~\bibnamefont {Chen}}, \bibinfo {author} {\bibfnamefont {C.}~\bibnamefont {Lee}}, \bibinfo {author} {\bibfnamefont {Y.}~\bibnamefont {Zhong}}, \bibinfo {author} {\bibfnamefont {S.}~\bibnamefont {Liu}}, \ and\ \bibinfo {author} {\bibfnamefont {D.}~\bibnamefont {Yu}},\ }\href {\doibase10.48550/arXiv.2303.04582} {\bibinfo {title} {Interaction-induced topological pumping in a solid-state quantum system},\ }\Eprint {https://arxiv.org/abs/2303.04582} {arxiv:2303.04582 [quant-ph]} \BibitemShut {NoStop}%
\bibitem [{\citenamefont {Jürgensen}\ \emph {et~al.}(2021)\citenamefont {Jürgensen}, \citenamefont {Mukherjee},\ and\ \citenamefont {Rechtsman}}]{jurgensenQuantizedNonlinearThouless2021}%
  \BibitemOpen
  \bibfield  {author} {\bibinfo {author} {\bibfnamefont {M.}~\bibnamefont {Jürgensen}}, \bibinfo {author} {\bibfnamefont {S.}~\bibnamefont {Mukherjee}}, \ and\ \bibinfo {author} {\bibfnamefont {M.~C.}\ \bibnamefont {Rechtsman}},\ }\bibfield  {title} {\bibinfo {title} {Quantized {{Nonlinear Thouless Pumping}}},\ }\href {\doibase10.1038/s41586-021-03688-9} {\bibfield  {journal} {\bibinfo  {journal} {Nature}\ }\textbf {\bibinfo {volume} {596}},\ \bibinfo {pages} {63} (\bibinfo {year} {2021})}\BibitemShut {NoStop}%
\bibitem [{\citenamefont {Fu}\ \emph {et~al.}(2022)\citenamefont {Fu}, \citenamefont {Wang}, \citenamefont {Kartashov}, \citenamefont {Konotop},\ and\ \citenamefont {Ye}}]{fuNonlinearThoulessPumping2022}%
  \BibitemOpen
  \bibfield  {author} {\bibinfo {author} {\bibfnamefont {Q.}~\bibnamefont {Fu}}, \bibinfo {author} {\bibfnamefont {P.}~\bibnamefont {Wang}}, \bibinfo {author} {\bibfnamefont {Y.~V.}\ \bibnamefont {Kartashov}}, \bibinfo {author} {\bibfnamefont {V.~V.}\ \bibnamefont {Konotop}}, \ and\ \bibinfo {author} {\bibfnamefont {F.}~\bibnamefont {Ye}},\ }\bibfield  {title} {\bibinfo {title} {Nonlinear {{Thouless Pumping}}: {{Solitons}} and {{Transport Breakdown}}},\ }\href {\doibase10.1103/PhysRevLett.128.154101} {\bibfield  {journal} {\bibinfo  {journal} {Phys. Rev. Lett.}\ }\textbf {\bibinfo {volume} {128}},\ \bibinfo {pages} {154101} (\bibinfo {year} {2022})}\BibitemShut {NoStop}%
\bibitem [{\citenamefont {Hayward}\ \emph {et~al.}(2018)\citenamefont {Hayward}, \citenamefont {Schweizer}, \citenamefont {Lohse}, \citenamefont {Aidelsburger},\ and\ \citenamefont {{Heidrich-Meisner}}}]{haywardTopologicalChargePumping2018}%
  \BibitemOpen
  \bibfield  {author} {\bibinfo {author} {\bibfnamefont {A.}~\bibnamefont {Hayward}}, \bibinfo {author} {\bibfnamefont {C.}~\bibnamefont {Schweizer}}, \bibinfo {author} {\bibfnamefont {M.}~\bibnamefont {Lohse}}, \bibinfo {author} {\bibfnamefont {M.}~\bibnamefont {Aidelsburger}}, \ and\ \bibinfo {author} {\bibfnamefont {F.}~\bibnamefont {{Heidrich-Meisner}}},\ }\bibfield  {title} {\bibinfo {title} {Topological charge pumping in the interacting bosonic {{Rice-Mele}} model},\ }\href {\doibase10.1103/PhysRevB.98.245148} {\bibfield  {journal} {\bibinfo  {journal} {Phys. Rev. B}\ }\textbf {\bibinfo {volume} {98}},\ \bibinfo {pages} {245148} (\bibinfo {year} {2018})}\BibitemShut {NoStop}%
\bibitem [{\citenamefont {Chong}\ \emph {et~al.}(2018)\citenamefont {Chong}, \citenamefont {Kim}, \citenamefont {Kim}, \citenamefont {Yoon}, \citenamefont {Kang},\ and\ \citenamefont {An}}]{chongObservationNonequilibriumSteady2018}%
  \BibitemOpen
  \bibfield  {author} {\bibinfo {author} {\bibfnamefont {K.~O.}\ \bibnamefont {Chong}}, \bibinfo {author} {\bibfnamefont {J.-R.}\ \bibnamefont {Kim}}, \bibinfo {author} {\bibfnamefont {J.}~\bibnamefont {Kim}}, \bibinfo {author} {\bibfnamefont {S.}~\bibnamefont {Yoon}}, \bibinfo {author} {\bibfnamefont {S.}~\bibnamefont {Kang}}, \ and\ \bibinfo {author} {\bibfnamefont {K.}~\bibnamefont {An}},\ }\bibfield  {title} {\bibinfo {title} {Observation of a non-equilibrium steady state of cold atoms in a moving optical lattice},\ }\href {\doibase10.1038/s42005-018-0024-5} {\bibfield  {journal} {\bibinfo  {journal} {Commun. Phys.}\ }\textbf {\bibinfo {volume} {1}},\ \bibinfo {pages} {1} (\bibinfo {year} {2018})}\BibitemShut {NoStop}%
\bibitem [{\citenamefont {Lin}\ \emph {et~al.}(2020)\citenamefont {Lin}, \citenamefont {Ke},\ and\ \citenamefont {Lee}}]{linInteractioninducedTopologicalBound2020}%
  \BibitemOpen
  \bibfield  {author} {\bibinfo {author} {\bibfnamefont {L.}~\bibnamefont {Lin}}, \bibinfo {author} {\bibfnamefont {Y.}~\bibnamefont {Ke}}, \ and\ \bibinfo {author} {\bibfnamefont {C.}~\bibnamefont {Lee}},\ }\bibfield  {title} {\bibinfo {title} {Interaction-induced topological bound states and {{Thouless}} pumping in a one-dimensional optical lattice},\ }\href {\doibase10.1103/PhysRevA.101.023620} {\bibfield  {journal} {\bibinfo  {journal} {Phys. Rev. A}\ }\textbf {\bibinfo {volume} {101}},\ \bibinfo {pages} {023620} (\bibinfo {year} {2020})}\BibitemShut {NoStop}%
\bibitem [{\citenamefont {{Arg{\"u}ello-Luengo}}\ \emph {et~al.}(2024)\citenamefont {{Arg{\"u}ello-Luengo}}, \citenamefont {Mark}, \citenamefont {Ferlaino}, \citenamefont {Lewenstein}, \citenamefont {Barbiero},\ and\ \citenamefont {{Juli{\`a}-Farr{\'e}}}}]{arguello-luengoStabilizationHubbardThoulessPumps2024}%
  \BibitemOpen
  \bibfield  {author} {\bibinfo {author} {\bibfnamefont {J.}~\bibnamefont {{Arg{\"u}ello-Luengo}}}, \bibinfo {author} {\bibfnamefont {M.~J.}\ \bibnamefont {Mark}}, \bibinfo {author} {\bibfnamefont {F.}~\bibnamefont {Ferlaino}}, \bibinfo {author} {\bibfnamefont {M.}~\bibnamefont {Lewenstein}}, \bibinfo {author} {\bibfnamefont {L.}~\bibnamefont {Barbiero}}, \ and\ \bibinfo {author} {\bibfnamefont {S.}~\bibnamefont {{Juli{\`a}-Farr{\'e}}}},\ }\bibfield  {title} {\bibinfo {title} {Stabilization of {{Hubbard-Thouless}} pumps through nonlocal fermionic repulsion},\ }\href {\doibase10.22331/q-2024-03-14-1285} {\bibfield  {journal} {\bibinfo  {journal} {Quantum}\ }\textbf {\bibinfo {volume} {8}},\ \bibinfo {pages} {1285} (\bibinfo {year} {2024})}\BibitemShut {NoStop}%
\bibitem [{\citenamefont {Walter}\ \emph {et~al.}(2023)\citenamefont {Walter}, \citenamefont {Zhu}, \citenamefont {Gächter}, \citenamefont {Minguzzi}, \citenamefont {Roschinski}, \citenamefont {Sandholzer}, \citenamefont {Viebahn},\ and\ \citenamefont {Esslinger}}]{walterQuantizationItsBreakdown2023}%
  \BibitemOpen
  \bibfield  {author} {\bibinfo {author} {\bibfnamefont {A.-S.}\ \bibnamefont {Walter}}, \bibinfo {author} {\bibfnamefont {Z.}~\bibnamefont {Zhu}}, \bibinfo {author} {\bibfnamefont {M.}~\bibnamefont {Gächter}}, \bibinfo {author} {\bibfnamefont {J.}~\bibnamefont {Minguzzi}}, \bibinfo {author} {\bibfnamefont {S.}~\bibnamefont {Roschinski}}, \bibinfo {author} {\bibfnamefont {K.}~\bibnamefont {Sandholzer}}, \bibinfo {author} {\bibfnamefont {K.}~\bibnamefont {Viebahn}}, \ and\ \bibinfo {author} {\bibfnamefont {T.}~\bibnamefont {Esslinger}},\ }\bibfield  {title} {\bibinfo {title} {Quantization and its breakdown in a {{Hubbard}}–{{Thouless}} pump},\ }\href {\doibase10.1038/s41567-023-02145-w} {\bibfield  {journal} {\bibinfo  {journal} {Nat. Phys.}\ }\textbf {\bibinfo {volume} {19}},\ \bibinfo {pages} {1471} (\bibinfo {year} {2023})}\BibitemShut {NoStop}%
\bibitem [{\citenamefont {Viebahn}\ \emph {et~al.}()\citenamefont {Viebahn}, \citenamefont {Walter}, \citenamefont {Bertok}, \citenamefont {Zhu}, \citenamefont {Gächter}, \citenamefont {Aligia}, \citenamefont {{Heidrich-Meisner}},\ and\ \citenamefont {Esslinger}}]{viebahnInteractioninducedChargePumping2023}%
  \BibitemOpen
  \bibfield  {author} {\bibinfo {author} {\bibfnamefont {K.}~\bibnamefont {Viebahn}}, \bibinfo {author} {\bibfnamefont {A.-S.}\ \bibnamefont {Walter}}, \bibinfo {author} {\bibfnamefont {E.}~\bibnamefont {Bertok}}, \bibinfo {author} {\bibfnamefont {Z.}~\bibnamefont {Zhu}}, \bibinfo {author} {\bibfnamefont {M.}~\bibnamefont {Gächter}}, \bibinfo {author} {\bibfnamefont {A.~A.}\ \bibnamefont {Aligia}}, \bibinfo {author} {\bibfnamefont {F.}~\bibnamefont {{Heidrich-Meisner}}}, \ and\ \bibinfo {author} {\bibfnamefont {T.}~\bibnamefont {Esslinger}},\ }\href {\doibase10.48550/arXiv.2308.03756} {\bibinfo {title} {Interaction-induced charge pumping in a topological many-body system},\ }\Eprint {https://arxiv.org/abs/2308.03756} {arxiv:2308.03756 [cond-mat, physics:physics]} \BibitemShut {NoStop}%
\bibitem [{\citenamefont {Zhu}\ \emph {et~al.}(2024)\citenamefont {Zhu}, \citenamefont {Gächter}, \citenamefont {Walter}, \citenamefont {Viebahn},\ and\ \citenamefont {Esslinger}}]{zhuReversalQuantizedHall2024}%
  \BibitemOpen
  \bibfield  {author} {\bibinfo {author} {\bibfnamefont {Z.}~\bibnamefont {Zhu}}, \bibinfo {author} {\bibfnamefont {M.}~\bibnamefont {Gächter}}, \bibinfo {author} {\bibfnamefont {A.-S.}\ \bibnamefont {Walter}}, \bibinfo {author} {\bibfnamefont {K.}~\bibnamefont {Viebahn}}, \ and\ \bibinfo {author} {\bibfnamefont {T.}~\bibnamefont {Esslinger}},\ }\bibfield  {title} {\bibinfo {title} {Reversal of quantized {{Hall}} drifts at noninteracting and interacting topological boundaries},\ }\href {\doibase10.1126/science.adg3848} {\bibfield  {journal} {\bibinfo  {journal} {Science}\ }\textbf {\bibinfo {volume} {384}},\ \bibinfo {pages} {317} (\bibinfo {year} {2024})}\BibitemShut {NoStop}%
\bibitem [{\citenamefont {Bertok}\ \emph {et~al.}(2022)\citenamefont {Bertok}, \citenamefont {{Heidrich-Meisner}},\ and\ \citenamefont {Aligia}}]{bertokSplittingTopologicalCharge2022}%
  \BibitemOpen
  \bibfield  {author} {\bibinfo {author} {\bibfnamefont {E.}~\bibnamefont {Bertok}}, \bibinfo {author} {\bibfnamefont {F.}~\bibnamefont {{Heidrich-Meisner}}}, \ and\ \bibinfo {author} {\bibfnamefont {A.~A.}\ \bibnamefont {Aligia}},\ }\bibfield  {title} {\bibinfo {title} {Splitting of topological charge pumping in an interacting two-component fermionic {{Rice-Mele Hubbard}} model},\ }\href {\doibase10.1103/PhysRevB.106.045141} {\bibfield  {journal} {\bibinfo  {journal} {Phys. Rev. B}\ }\textbf {\bibinfo {volume} {106}},\ \bibinfo {pages} {045141} (\bibinfo {year} {2022})}\BibitemShut {NoStop}%
\bibitem [{\citenamefont {Rachel}(2018)}]{rachelInteractingTopologicalInsulators2018}%
  \BibitemOpen
  \bibfield  {author} {\bibinfo {author} {\bibfnamefont {S.}~\bibnamefont {Rachel}},\ }\bibfield  {title} {\bibinfo {title} {Interacting topological insulators: A review},\ }\href {\doibase10.1088/1361-6633/aad6a6} {\bibfield  {journal} {\bibinfo  {journal} {Rep. Prog. Phys.}\ }\textbf {\bibinfo {volume} {81}},\ \bibinfo {pages} {116501} (\bibinfo {year} {2018})}\BibitemShut {NoStop}%
\bibitem [{\citenamefont {Wang}\ \emph {et~al.}(2014)\citenamefont {Wang}, \citenamefont {Potter},\ and\ \citenamefont {Senthil}}]{wangClassificationInteractingElectronic2014}%
  \BibitemOpen
  \bibfield  {author} {\bibinfo {author} {\bibfnamefont {C.}~\bibnamefont {Wang}}, \bibinfo {author} {\bibfnamefont {A.~C.}\ \bibnamefont {Potter}}, \ and\ \bibinfo {author} {\bibfnamefont {T.}~\bibnamefont {Senthil}},\ }\bibfield  {title} {\bibinfo {title} {Classification of {{Interacting Electronic Topological Insulators}} in {{Three Dimensions}}},\ }\href {\doibase10.1126/science.1243326} {\bibfield  {journal} {\bibinfo  {journal} {Science}\ }\textbf {\bibinfo {volume} {343}},\ \bibinfo {pages} {629} (\bibinfo {year} {2014})}\BibitemShut {NoStop}%
\bibitem [{\citenamefont {Jünemann}\ \emph {et~al.}(2017)\citenamefont {Jünemann}, \citenamefont {Piga}, \citenamefont {Ran}, \citenamefont {Lewenstein}, \citenamefont {Rizzi},\ and\ \citenamefont {Bermudez}}]{junemannExploringInteractingTopological2017}%
  \BibitemOpen
  \bibfield  {author} {\bibinfo {author} {\bibfnamefont {J.}~\bibnamefont {Jünemann}}, \bibinfo {author} {\bibfnamefont {A.}~\bibnamefont {Piga}}, \bibinfo {author} {\bibfnamefont {S.-J.}\ \bibnamefont {Ran}}, \bibinfo {author} {\bibfnamefont {M.}~\bibnamefont {Lewenstein}}, \bibinfo {author} {\bibfnamefont {M.}~\bibnamefont {Rizzi}}, \ and\ \bibinfo {author} {\bibfnamefont {A.}~\bibnamefont {Bermudez}},\ }\bibfield  {title} {\bibinfo {title} {Exploring {{Interacting Topological Insulators}} with {{Ultracold Atoms}}: {{The Synthetic Creutz-Hubbard Model}}},\ }\href {\doibase10.1103/PhysRevX.7.031057} {\bibfield  {journal} {\bibinfo  {journal} {Phys. Rev. X}\ }\textbf {\bibinfo {volume} {7}},\ \bibinfo {pages} {031057} (\bibinfo {year} {2017})}\BibitemShut {NoStop}%
\bibitem [{\citenamefont {{de Léséleuc}}\ \emph {et~al.}(2019)\citenamefont {{de Léséleuc}}, \citenamefont {Lienhard}, \citenamefont {Scholl}, \citenamefont {Barredo}, \citenamefont {Weber}, \citenamefont {Lang}, \citenamefont {Büchler}, \citenamefont {Lahaye},\ and\ \citenamefont {Browaeys}}]{deleseleucObservationSymmetryprotectedTopological2019}%
  \BibitemOpen
  \bibfield  {author} {\bibinfo {author} {\bibfnamefont {S.}~\bibnamefont {{de Léséleuc}}}, \bibinfo {author} {\bibfnamefont {V.}~\bibnamefont {Lienhard}}, \bibinfo {author} {\bibfnamefont {P.}~\bibnamefont {Scholl}}, \bibinfo {author} {\bibfnamefont {D.}~\bibnamefont {Barredo}}, \bibinfo {author} {\bibfnamefont {S.}~\bibnamefont {Weber}}, \bibinfo {author} {\bibfnamefont {N.}~\bibnamefont {Lang}}, \bibinfo {author} {\bibfnamefont {H.~P.}\ \bibnamefont {Büchler}}, \bibinfo {author} {\bibfnamefont {T.}~\bibnamefont {Lahaye}}, \ and\ \bibinfo {author} {\bibfnamefont {A.}~\bibnamefont {Browaeys}},\ }\bibfield  {title} {\bibinfo {title} {Observation of a symmetry-protected topological phase of interacting bosons with {{Rydberg}} atoms},\ }\href {\doibase10.1126/science.aav9105} {\bibfield  {journal} {\bibinfo  {journal} {Science}\ }\textbf {\bibinfo {volume} {365}},\ \bibinfo {pages} {775} (\bibinfo {year} {2019})}\BibitemShut {NoStop}%
\bibitem [{\citenamefont {Lu}\ and\ \citenamefont {Vishwanath}(2012)}]{luTheoryClassificationInteracting2012}%
  \BibitemOpen
  \bibfield  {author} {\bibinfo {author} {\bibfnamefont {Y.-M.}\ \bibnamefont {Lu}}\ and\ \bibinfo {author} {\bibfnamefont {A.}~\bibnamefont {Vishwanath}},\ }\bibfield  {title} {\bibinfo {title} {Theory and classification of interacting integer topological phases in two dimensions: {{A Chern-Simons}} approach},\ }\href {\doibase10.1103/PhysRevB.86.125119} {\bibfield  {journal} {\bibinfo  {journal} {Phys. Rev. B}\ }\textbf {\bibinfo {volume} {86}},\ \bibinfo {pages} {125119} (\bibinfo {year} {2012})}\BibitemShut {NoStop}%
\bibitem [{\citenamefont {Freedman}\ \emph {et~al.}(2004)\citenamefont {Freedman}, \citenamefont {Nayak}, \citenamefont {Shtengel}, \citenamefont {Walker},\ and\ \citenamefont {Wang}}]{freedmanClassTinvariantTopological2004}%
  \BibitemOpen
  \bibfield  {author} {\bibinfo {author} {\bibfnamefont {M.}~\bibnamefont {Freedman}}, \bibinfo {author} {\bibfnamefont {C.}~\bibnamefont {Nayak}}, \bibinfo {author} {\bibfnamefont {K.}~\bibnamefont {Shtengel}}, \bibinfo {author} {\bibfnamefont {K.}~\bibnamefont {Walker}}, \ and\ \bibinfo {author} {\bibfnamefont {Z.}~\bibnamefont {Wang}},\ }\bibfield  {title} {\bibinfo {title} {A class of {{P}},{{T-invariant}} topological phases of interacting electrons},\ }\href {\doibase10.1016/j.aop.2004.01.006} {\bibfield  {journal} {\bibinfo  {journal} {Ann. Phys. (N. Y.)}\ }\textbf {\bibinfo {volume} {310}},\ \bibinfo {pages} {428} (\bibinfo {year} {2004})}\BibitemShut {NoStop}%
\bibitem [{\citenamefont {Ke}\ \emph {et~al.}(2017)\citenamefont {Ke}, \citenamefont {Qin}, \citenamefont {Kivshar},\ and\ \citenamefont {Lee}}]{keMultiparticleWannierStates2017}%
  \BibitemOpen
  \bibfield  {author} {\bibinfo {author} {\bibfnamefont {Y.}~\bibnamefont {Ke}}, \bibinfo {author} {\bibfnamefont {X.}~\bibnamefont {Qin}}, \bibinfo {author} {\bibfnamefont {Y.~S.}\ \bibnamefont {Kivshar}}, \ and\ \bibinfo {author} {\bibfnamefont {C.}~\bibnamefont {Lee}},\ }\bibfield  {title} {\bibinfo {title} {Multiparticle {{Wannier}} states and {{Thouless}} pumping of interacting bosons},\ }\href {\doibase10.1103/PhysRevA.95.063630} {\bibfield  {journal} {\bibinfo  {journal} {Phys. Rev. A}\ }\textbf {\bibinfo {volume} {95}},\ \bibinfo {pages} {063630} (\bibinfo {year} {2017})}\BibitemShut {NoStop}%
\bibitem [{\citenamefont {Zhang}\ and\ \citenamefont {Liu}(2022)}]{zhangUnconventionalFloquetTopological2022}%
  \BibitemOpen
  \bibfield  {author} {\bibinfo {author} {\bibfnamefont {L.}~\bibnamefont {Zhang}}\ and\ \bibinfo {author} {\bibfnamefont {X.-J.}\ \bibnamefont {Liu}},\ }\bibfield  {title} {\bibinfo {title} {Unconventional {{Floquet Topological Phases}} from {{Quantum Engineering}} of {{Band-Inversion Surfaces}}},\ }\href {\doibase10.1103/PRXQuantum.3.040312} {\bibfield  {journal} {\bibinfo  {journal} {PRX Quantum}\ }\textbf {\bibinfo {volume} {3}},\ \bibinfo {pages} {040312} (\bibinfo {year} {2022})}\BibitemShut {NoStop}%
\bibitem [{\citenamefont {Yan}\ and\ \citenamefont {Zhou}(2018)}]{yanYangMonopolesEmergent2018}%
  \BibitemOpen
  \bibfield  {author} {\bibinfo {author} {\bibfnamefont {Y.}~\bibnamefont {Yan}}\ and\ \bibinfo {author} {\bibfnamefont {Q.}~\bibnamefont {Zhou}},\ }\bibfield  {title} {\bibinfo {title} {Yang {{Monopoles}} and {{Emergent Three-Dimensional Topological Defects}} in {{Interacting Bosons}}},\ }\href {\doibase10.1103/PhysRevLett.120.235302} {\bibfield  {journal} {\bibinfo  {journal} {Phys. Rev. Lett.}\ }\textbf {\bibinfo {volume} {120}},\ \bibinfo {pages} {235302} (\bibinfo {year} {2018})}\BibitemShut {NoStop}%
\bibitem [{sup()}]{supnote}%
  \BibitemOpen
  \href@noop {} {}\bibinfo {note} {The supplemental material at TO.BE.INSERT.BY.THE.EDITOR contains more simulation results under different parameter settings, derivation of the effective Hamiltonian of the bosonic and SU(N) fermionic system, the comparison of the effective Hamiltonian with the exact diagonalization result, discussion of the mean-field limit and the discussion of initial state preparation.}\BibitemShut {Stop}%
\bibitem [{\citenamefont {Fishman}\ \emph {et~al.}(2022)\citenamefont {Fishman}, \citenamefont {White},\ and\ \citenamefont {Stoudenmire}}]{fishmanITensorSoftwareLibrary2022}%
  \BibitemOpen
  \bibfield  {author} {\bibinfo {author} {\bibfnamefont {M.}~\bibnamefont {Fishman}}, \bibinfo {author} {\bibfnamefont {S.}~\bibnamefont {White}}, \ and\ \bibinfo {author} {\bibfnamefont {E.}~\bibnamefont {Stoudenmire}},\ }\bibfield  {title} {\bibinfo {title} {The {{ITensor Software Library}} for {{Tensor Network Calculations}}},\ }\href {\doibase10.21468/SciPostPhysCodeb.4} {\bibfield  {journal} {\bibinfo  {journal} {SciPost Physics Codebases}\ ,\ \bibinfo {pages} {004}} (\bibinfo {year} {2022})}\BibitemShut {NoStop}%
\bibitem [{\citenamefont {Liu}\ \emph {et~al.}(2023)\citenamefont {Liu}, \citenamefont {Hu}, \citenamefont {Zhang}, \citenamefont {Ke},\ and\ \citenamefont {Lee}}]{liuCorrelatedTopologicalPumping2023}%
  \BibitemOpen
  \bibfield  {author} {\bibinfo {author} {\bibfnamefont {W.}~\bibnamefont {Liu}}, \bibinfo {author} {\bibfnamefont {S.}~\bibnamefont {Hu}}, \bibinfo {author} {\bibfnamefont {L.}~\bibnamefont {Zhang}}, \bibinfo {author} {\bibfnamefont {Y.}~\bibnamefont {Ke}}, \ and\ \bibinfo {author} {\bibfnamefont {C.}~\bibnamefont {Lee}},\ }\bibfield  {title} {\bibinfo {title} {Correlated topological pumping of interacting bosons assisted by {{Bloch}} oscillations},\ }\href {\doibase10.1103/PhysRevResearch.5.013020} {\bibfield  {journal} {\bibinfo  {journal} {Phys. Rev. Res.}\ }\textbf {\bibinfo {volume} {5}},\ \bibinfo {pages} {013020} (\bibinfo {year} {2023})}\BibitemShut {NoStop}%
\bibitem [{\citenamefont {Wannier}(1960)}]{wannierWaveFunctionsEffective1960}%
  \BibitemOpen
  \bibfield  {author} {\bibinfo {author} {\bibfnamefont {G.~H.}\ \bibnamefont {Wannier}},\ }\bibfield  {title} {\bibinfo {title} {Wave {{Functions}} and {{Effective Hamiltonian}} for {{Bloch Electrons}} in an {{Electric Field}}},\ }\href {\doibase10.1103/PhysRev.117.432} {\bibfield  {journal} {\bibinfo  {journal} {Phys. Rev.}\ }\textbf {\bibinfo {volume} {117}},\ \bibinfo {pages} {432} (\bibinfo {year} {1960})}\BibitemShut {NoStop}%
\bibitem [{\citenamefont {Hartmann}\ \emph {et~al.}(2004)\citenamefont {Hartmann}, \citenamefont {Keck}, \citenamefont {Korsch},\ and\ \citenamefont {Mossmann}}]{hartmannDynamicsBlochOscillations2004}%
  \BibitemOpen
  \bibfield  {author} {\bibinfo {author} {\bibfnamefont {T.}~\bibnamefont {Hartmann}}, \bibinfo {author} {\bibfnamefont {F.}~\bibnamefont {Keck}}, \bibinfo {author} {\bibfnamefont {H.~J.}\ \bibnamefont {Korsch}}, \ and\ \bibinfo {author} {\bibfnamefont {S.}~\bibnamefont {Mossmann}},\ }\bibfield  {title} {\bibinfo {title} {Dynamics of {{Bloch}} oscillations},\ }\href {\doibase10.1088/1367-2630/6/1/002} {\bibfield  {journal} {\bibinfo  {journal} {New J. Phys.}\ }\textbf {\bibinfo {volume} {6}},\ \bibinfo {pages} {2} (\bibinfo {year} {2004})}\BibitemShut {NoStop}%
\bibitem [{\citenamefont {Holthaus}\ and\ \citenamefont {Hone}(1996)}]{holthausLocalizationEffectsAcdriven1996}%
  \BibitemOpen
  \bibfield  {author} {\bibinfo {author} {\bibfnamefont {M.}~\bibnamefont {Holthaus}}\ and\ \bibinfo {author} {\bibfnamefont {D.~W.}\ \bibnamefont {Hone}},\ }\bibfield  {title} {\bibinfo {title} {Localization effects in ac-driven tight-binding lattices},\ }\href {\doibase10.1080/01418639608240331} {\bibfield  {journal} {\bibinfo  {journal} {Philosophical Magazine B}\ }\textbf {\bibinfo {volume} {74}},\ \bibinfo {pages} {105} (\bibinfo {year} {1996})}\BibitemShut {NoStop}%
\bibitem [{\citenamefont {Dunlap}\ and\ \citenamefont {Kenkre}(1986)}]{dunlapDynamicLocalizationCharged1986}%
  \BibitemOpen
  \bibfield  {author} {\bibinfo {author} {\bibfnamefont {D.~H.}\ \bibnamefont {Dunlap}}\ and\ \bibinfo {author} {\bibfnamefont {V.~M.}\ \bibnamefont {Kenkre}},\ }\bibfield  {title} {\bibinfo {title} {Dynamic localization of a charged particle moving under the influence of an electric field},\ }\href {\doibase10.1103/PhysRevB.34.3625} {\bibfield  {journal} {\bibinfo  {journal} {Phys. Rev. B}\ }\textbf {\bibinfo {volume} {34}},\ \bibinfo {pages} {3625} (\bibinfo {year} {1986})}\BibitemShut {NoStop}%
\bibitem [{\citenamefont {Ho}\ and\ \citenamefont {Li}(2017)}]{hoChernNumbersInteractionstretched2017}%
  \BibitemOpen
  \bibfield  {author} {\bibinfo {author} {\bibfnamefont {T.-L.}\ \bibnamefont {Ho}}\ and\ \bibinfo {author} {\bibfnamefont {C.}~\bibnamefont {Li}},\ }\href {\doibase10.48550/arXiv.1704.03833} {\bibinfo {title} {The {{Chern Numbers}} of {{Interaction-stretched Monopoles}} in {{Spinor Bose Condensates}}},\ } (\bibinfo {year} {2017}),\ \Eprint {https://arxiv.org/abs/1704.03833} {arxiv:1704.03833 [cond-mat]} \BibitemShut {NoStop}%
\end{thebibliography}%

%%%%%%%%%% Merge with supplemental materials %%%%%%%%%%
\clearpage
\widetext
\begin{center}
\textbf{\large Supplemental Material: Interaction induced splitting of Dirac monopoles in the topological Thouless pumping of strongly interacting Bosons and SU($N$) Fermions}
\end{center}
%%%%%%%%%% Merge with supplemental materials %%%%%%%%%%
%%%%%%%%%% Prefix a "S" to all equations, figures, tables and reset the counter %%%%%%%%%%
\setcounter{equation}{0}
\setcounter{figure}{0}
\setcounter{table}{0}
\setcounter{page}{8}
\makeatletter
\renewcommand{\theequation}{S\arabic{equation}}
\renewcommand{\thefigure}{S\arabic{figure}}
\renewcommand{\bibnumfmt}[1]{[S#1]}
% \renewcommand{\thesection}{\arabic{section}}
% \renewcommand \thesection{S\@arabic\c@section}
% \renewcommand{\citenumfont}[1]{S#1}
%%%%%%%%%% Prefix a "S" to all equations, figures, tables and reset the counter %%%%%%%%%%

\tableofcontents

% ****** Start of file apssamp.tex ******
%
%   This file is part of the APS files in the REVTeX 4.2 distribution.
%   Version 4.2a of REVTeX, December 2014
%
%   Copyright (c) 2014 The American Physical Society.
%

%   See the REVTeX 4 README file for restrictions and more information.
% 
% TeX'ing this file requires that you have AMS-LaTeX 2.0 installed
% as well as the rest of the prerequisites for REVTeX 4.2
%
% See the REVTeX 4 README file
% It also requires running BibTeX. The commands are as follows:
%
%  1)  latex apssamp.tex
%  2)  bibtex apssamp
%  3)  latex apssamp.tex
%  4)  latex apssamp.texf
%
% \documentclass[
%  %twocolumn,
%  %prl,
% superscriptaddress,
% %groupedaddress,
% %unsortedaddress,
% %runinaddress,
% %frontmatterverbose, 
% %preprint,
% %preprintnumbers,
% %nofootinbib,
% %nobibnotes,
% %bibnotes,
%  amsmath,amssymb,
%  aps,
% pra,
% %prb,
% %rmp,
% %prstab,
% %prstper,
% floatfix,
% ]{revtex4-2}

% \usepackage{graphicx}% Include figure files
% \usepackage{dcolumn}% Align table columns on decimal point
% \usepackage{bm}% bold math
% \usepackage{grffile} 
% \usepackage{subfigure}
% \usepackage[marginal]{footmisc}
% \usepackage{epstopdf}
% \usepackage{amssymb}
% \usepackage{mathrsfs}
% \usepackage{braket}
% \usepackage{amsmath,mathtools,amsfonts,bm,graphicx}
% \usepackage{upgreek}
% \usepackage{color}
% \usepackage{etoolbox}
% \usepackage[colorlinks=true,linkcolor=blue,urlcolor=blue,citecolor=blue]{hyperref} % conflict with arXiv
% \usepackage{multirow}
% \usepackage[normalem]{ulem}
% \begin{document}

\preprint{APS/123-QED}

\title{Supplemental Material: Interaction induced splitting of Dirac monopoles in the topological Thouless pumping of strongly interacting Bosons and SU($N$) Fermions}% Force line breaks with \\

\author{Lam Hei}
\affiliation{%
Department of Physics, The Chinese University of Hong Kong, Shatin, New Territories, Hong Kong, China
}%

\author{Yangqian Yan}%
 \email{yqyan@cuhk.edu.hk}
\affiliation{%
Department of Physics, The Chinese University of Hong Kong, Shatin, New Territories, Hong Kong, China
}
\affiliation{
The Chinese University of Hong Kong Shenzhen Research Institute, 518057 Shenzhen, China
}%

\date{\today}% It is always \today, today,

\maketitle

\tableofcontents

\section{Simulation of the adiabatic evolution of Thouless pumping in interacting bosonic and fermionic model }

\subsection{Simulation of Bosonic Pumping}

\begin{figure}[h]
    \centering
    \includegraphics[scale=1.1]{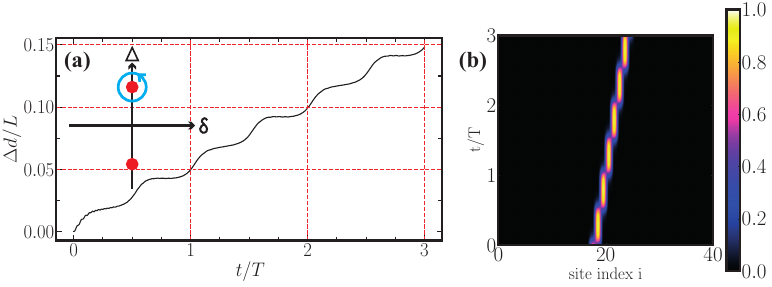}
    \caption{Simulation of the adiabatic evolution of the pumping by placing a single boson in the middle of the system on the top of a half-filled lattice (a) The relative shift of the center of mass of the single boson (b). Heat Map of the evolution of the single boson during the adiabatic evolution.}
    \label{fig:boson_np1}
\end{figure}

In the simulation, we employ the Time-Evolving Block Decimation (TEBD) technique to investigate the time dynamics of adiabatic particle pumping. A single boson is placed in the center of a lattice with a background of 1 particle filling per unit cell. Our computational lattice is defined with a system size of \(L = 40\), interaction strength \(U = 30\), and tunneling energy \(TJ = 30\), where \(J = 1\) is the hopping parameter. The on-site potential strength is set to \(\Delta_{0} = 5\). A tilted potential field is introduced to localize the wavefunction and prevent its dispersion.

At \(t = 0\), the system is initialized in the dimerized phase, characterized by the difference in hopping amplitudes \(\delta = J_{2} - J_{1} = 2J_{0}\), with the inter-cell hopping effectively reduced to zero, ensuring that the system resides in an insulating phase. The iTEBD simulation is then used to prepare the initial state corresponding to the system's ground state.

The effective Hamiltonian determines the monopole’s position. As depicted in Fig.~\ref{fig:boson_np1}(a), we conduct three complete pumping cycles to circulate the monopole characterizing the pumping of this single boson. In each cycle, the particle undergoes a complete quantized pumping process. After three cycles, the system’s center of mass exhibits a relative shift of \(3/L\), corresponding to the precise three unit cells displacement expected for the single-particle case.

The close correspondence between our simulation results and the effective Hamiltonian predictions corroborates our model’s validity.

\subsection{Simulation of Bosons system with uniform filling along different trajectories in the parameter space }

\begin{figure}[h]
    \centering
    \includegraphics[scale=1.0]{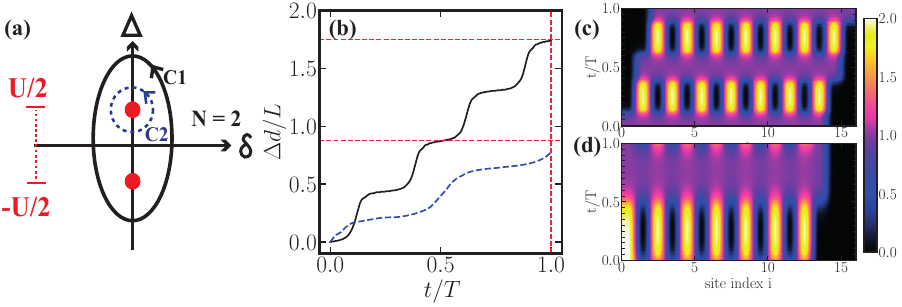}
    \caption{The simulation of an evenly filled system where the last unit cells are left empty to avoid the boundary effect. (a) The schematic diagram of different trajectories chosen in the simulation (b)  The relative shift of the center of mass along different trajectories. (c) The heat map of the particle density during the evolution along the trajectory of C1 and (d) C2.}
    \label{fig:boson_2p_even}
\end{figure}

\subsection {Simulation of the SU(2) model with uniform occupation number in different parameter setting  }

\begin{figure}[h]
    \centering
    \includegraphics[scale=0.8]{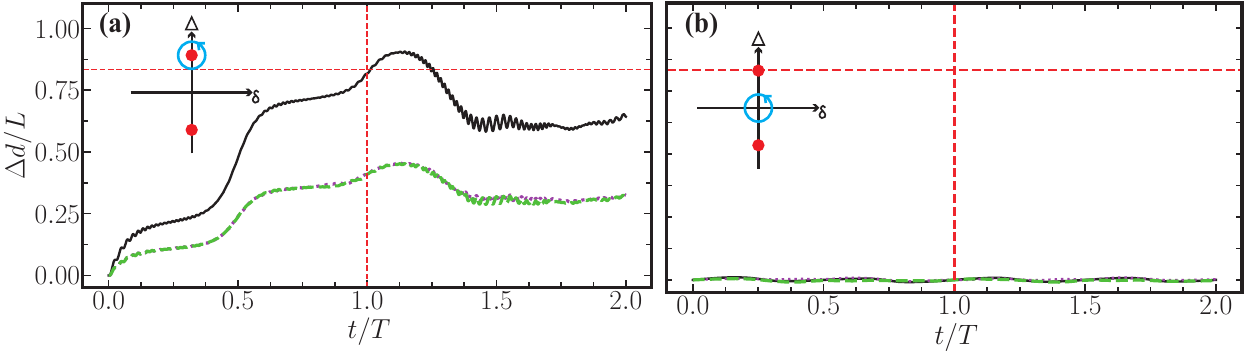}
    \caption{The shift of the center of mass of the SU(2) model with a fully filled two particles in every unit cell in different locations of $\Delta-\delta$ space. The black solid line corresponds to the total particle pumping during the evolution, while the green dashed line (purpled dotted line ) corresponds to the spin-up (spin-down) particle pumping. (a) trajectory run along the path $(\Delta,\delta) = (\Delta(t) + U/2, 2J_{0}\text{cos}(2\pi t/T) )$ (b) $(\Delta,\delta) = (\Delta(t), 2J_{0}\text{cos}(2\pi t/T) )$  }
    \label{fig:Su2-2l}
\end{figure}

In our study of the SU(2) fermionic system, simulations were performed along two distinct trajectories for a system of size \(L = 12\), employing the same system parameters as those used in the bosonic case. The initial state was prepared using the iTEBD algorithm, with each unit cell in the system uniformly populated by one spin-up and one spin-down fermion.

As illustrated in Fig.~\ref{fig:Su2-2l}(a), quantized pumping is evident during the first cycle. However, this quantization breaks down in subsequent cycles due to the band degeneracy issue arising from the permutation of spin particles during the evolution. When the trajectory encircles the origin, as shown in Fig.~\ref{fig:Su2-2l}(b), particle pumping is not observed even in the first cycle. This phenomenon is attributed to the even-odd particle filling issue, similar to the bosonic case; hence, the presence of a monopole at the origin should not be anticipated for a system with two-particle filling.

\subsection{SU(2) model with and without staggered field}
\begin{figure}[h]
    \centering
    \includegraphics[scale=0.95]{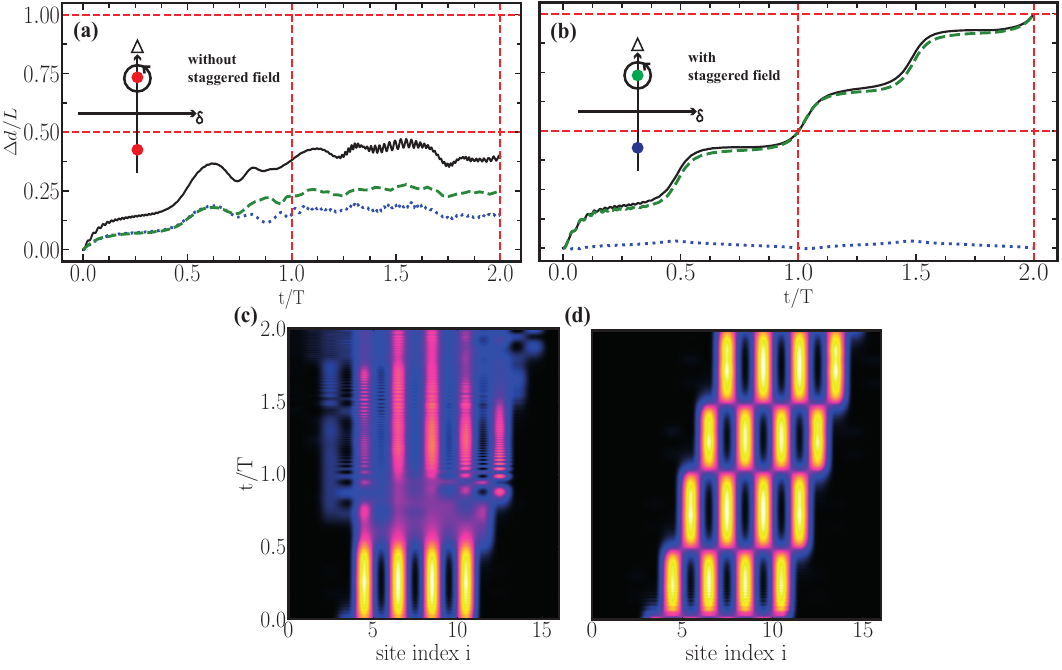}
    \caption{ The shift of the center of mass of the SU(2) model. 4 spin-up particles are placed in the center of the system in the background of 1 spin-down particle filling per unit-cell  (a) without the background staggered field, (b) and the addition of staggered field coupling to 2 different spin component (c) The heat map of all the spin-up particle density in the system during the evolution in the background of without staggered field (d) with staggered field. }
    \label{fig:Su2woB_vs_B}
\end{figure}
Consider a system where each unit cell is uniformly filled with a spin-up particle and half the number of spin-down particles are placed at the system’s center. Following our discussion on the degeneracy of the ground state band, which arises from permuting different spin species, we propose that at \(t = \frac{3T}{4}\), the concept of a Dirac monopole becomes ambiguous, and quantized pumping is not expected, as depicted in Fig.~\ref{fig:Su2woB_vs_B}(a). Nevertheless, by introducing a Zeeman field, the degeneracy between the two lowest bands is lifted, effectively recovering two non-degenerate energy bands and restoring quantized pumping, as demonstrated in Fig.~\ref{fig:Su2woB_vs_B}(b). The coupling strength of the Zeeman field is set to \(BJ = 2\), and the energy coupling to the staggered Zeeman field for the two different spin species is given by \(B\sigma_{z}\).

\subsection{Charge pumping in SU(3) model }

\begin{figure}
    \centering
    \includegraphics[scale=1]{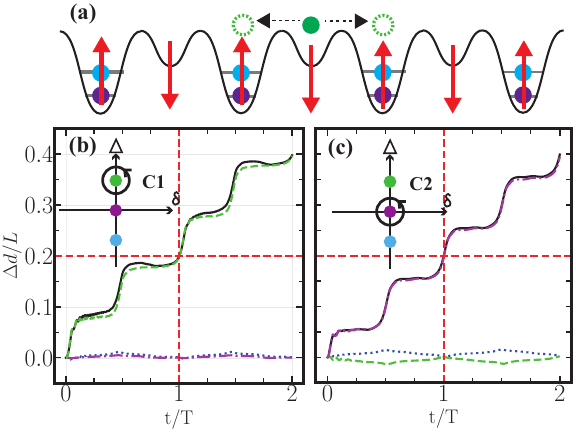}
    \caption{(a). Schematic diagram of the particle pumping of an SU(3) system around the C1 trajectory with two different flavors used to fill up all the unit cells, and a third type of fermion is placed independently in the system's center. (b),(c).The relative shift of the center of mass of different fermion species during the system evolution along the trajectories of C1, C2 }
    \label{fig:su3_pump}
\end{figure}

To further demonstrate the validity of our theory, we performed numerical simulations on an SU(3) symmetric model along different trajectories by introducing a single fermion onto a filled system composed of two other types of fermions. The simulation was executed in a system with a size of $L = 12$. The parameters were set to $J_{0} = 1$, $\Delta_{0} = 5$, $U=30$, and $B = 2$. Before the simulation, the iTEBD algorithm was used to find the ground state within the dimerized phase.

As depicted in Fig.~\ref{fig:su3_pump}(b) and (c), two trajectories, C1 and C2, were simulated to encircle the monopoles distinguished by different types of fermions. Quantized pumping only occurred for the particular type of fermion species depending on the choice of trajectory encircling different monopoles described by different effective Hamiltonians. Due to the effect of even-odd particle filling, a monopole is located at the origin, contributing to extra-particle pumping. These results are similar to what we obtained in the bosonic case.

\section{Effective Hamiltonian Construction for Bosonic System  }
\subsection{Effective Hamiltonian of the single particle pumping}

Selecting the basis from the subspace of the original unperturbed full Hamiltonian, the basis set is expressed as $\hat{a}^{\dagger}_{i}\ket{\text{G.S.}}$ and $\hat{b}^{\dagger}_{i}\ket{\text{G.S.}}$, where $\ket{\text{G.S.}} = \prod_{i} (\hat{a}^{\dagger}_{i})^{q}(\hat{b}_{i}^{\dagger})^{N-q}\ket{0}$. Consequently, we construct the new Hamiltonian in real space from degenerate perturbation theory by treating the hopping term as a small perturbation. The new Hamiltonian is written as:
\begin{equation}
    \hat{H}_{\text{eff}} = \sum_{i=1}^{L} R_{1}(t) \hat{a}_{i}^{\dagger}\hat{b}_{i+1} + R_{2}(t)\hat{b}_{i}^{\dagger}\hat{a}_{i+1} + \text{h.c.} + \Delta'(t)\hat{n}_{a,i} - \Delta'(t)\hat{n}_{b,i} + H_{0},
    \label{eq:Heff}
\end{equation}
where
\begin{equation}
    \begin{split}
        H_{0} & = (L-1)\left(-q\Delta + (N-q)\Delta + \frac{U}{2}q(q-1) + \frac{U}{2}(N-q)(N-q-1)\right) \\
        & \quad -q\Delta + (N-q)\Delta + \frac{U}{2}\left( q^{2} + (N-q)^{2}\right), \\
        R_{1}(t) & = \sqrt{N-q+1}\sqrt{q+1}J_{0}\left[1-\cos\left(\frac{2\pi t}{T}\right)\right], \\
        R_{2}(t) & = \sqrt{N-q+1}\sqrt{q+1}J_{0}\left[1+\cos\left(\frac{2\pi t}{T}\right)\right], \\
        \Delta'(t) & = \Delta(t) + \frac{U}{2}\left(q - (N-q)\right).
    \end{split}
    \label{eq:H0R1R2Delta}
\end{equation}
Subsequently, we perform a Fourier transform to express the effective Hamiltonian in quasi-momentum space. We find that it resembles a non-interacting Rice-Mele (RM) model with re-parameterized parameters in the $B$-parameter space, $\vec{B}\cdot\vec{\sigma}$. The two-band Hamiltonian is represented as follows:
\begin{equation}
    H(k) = H_{0}\sigma_{0} + \left( R_{1}(t) + R_{2}(t)\cos(k) \right) \sigma_{x} + R_{2}(t)\sin(k)\sigma_{y} + \Delta'(t)\sigma_{z}.
    \label{eq:Hk}
\end{equation}
Solving the energy spectrum, the Dirac monopole can be easily located along the $\Delta$ axis in the parameter space which depends on the interaction strength and the particle filling. The spacing of every nearest Dirac monopole along the $\Delta$ axis is given by $\Delta_{n} - \Delta_{n+1} = U$.

\subsection{Effective Hamiltonian of the single hole pumping}
We performed calculations on a system comprising two particles uniformly distributed in each unit cell, except one unit cell occupied by a single particle. This defect within the unit cell can be conceptualized as a `hole' created by annihilating a specific particle in the system's ground state. We anticipated that the topological defect, characterizing the pumping of the hole, would manifest at the origin of the parameter space—a scenario analogous to an $N+1$ layer system, but notably in the context of `hole' creation.
To derive the effective Hamiltonian, we followed a similar procedure as before, but with our Hamiltonian constructed using the basis $\{\hat{a}_{i}\ket{G.S.}, \hat{b}_{i}\ket{G.S.}\}$, where $i \in [1,L]$. Consequently, the effective Hamiltonian is given by:
\begin{equation}
H_{\text{eff}} = 
\begin{pmatrix}
\Delta - \frac{U}{2}(N-2q) & R_{1}+R_{2}e^{\mathbf{{i}}k} \\
R_{1}+R_{2}e^{-\mathbf{i}k} & -\Delta + \frac{U}{2}(N-2q) 
\end{pmatrix},
\end{equation}
As indicated above, the effective Hamiltonian resulting from the system with a hole is related to the effective Hamiltonian of a single particle in an $N+1$ layer system via the inversion operator $\sigma_{y}$. The energy bands are inverted, with the upper band becoming the lower band and vice versa. Therefore, the lowest energy states are expected to exhibit the opposite Chern number compared to the $N+1$ layer system.
Subsequent simulations involved tracing a trajectory around the origin in parameter space, and the quantized pumping of the hole was observed. The holes were pumped in the opposite direction, which is equivalent to shifting the center of mass in the reverse direction.

\section{Comparison of the energy spectrum of the Excat Diangonlization and the effective Hamiltonian}
Leveraging the concept of multiparticle Wannier states as described by Ke et al.~\cite{keMultiparticleWannierStates2017}, we directly diagonalize the original Hamiltonian in the quasi-momentum space. We then compare these results with those derived from the effective Hamiltonian. This comparison was conducted in a system comprising four unit cells, with five particles populating the entire lattice. The calculation is carried out at the moment $t = T / 4$ with the strength of the interaction $U = 30, J = 1$, and $\Delta = 5$.  

\begin{figure}[h]
    \centering
    \includegraphics[scale=1.05]{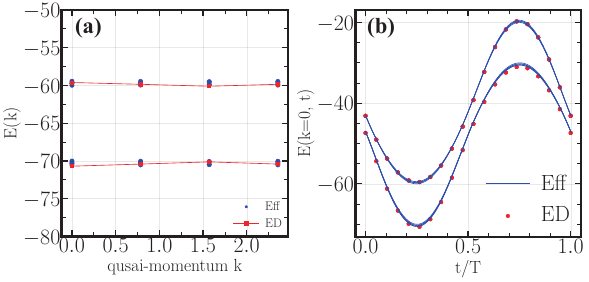}
    \caption{Comparison of the energy spectrum of the interacting bosonic model with the exact diagonalization method and the effective Hamiltonian (a) The energy spectrum of the system at $t = T/4$ (b) The time dynamic of the system at k = 0.}
    \label{fig:s1}
\end{figure}

As depicted in Fig.~\ref{fig:s1}, the effective Hamiltonian's energy spectrum closely mirrors the original Hamiltonian's lowest two energy bands within the relevant parameter space. Upon examining the wavefunctions corresponding to the lowest two bands, it becomes evident that they align with the states utilized to formulate the effective Hamiltonian during the perturbation analysis.

\section{Calculation of the Mean-field approximation in the interacting bosonic Hamiltonian}
 \begin{figure}
    \centering
    \includegraphics[scale=0.95]{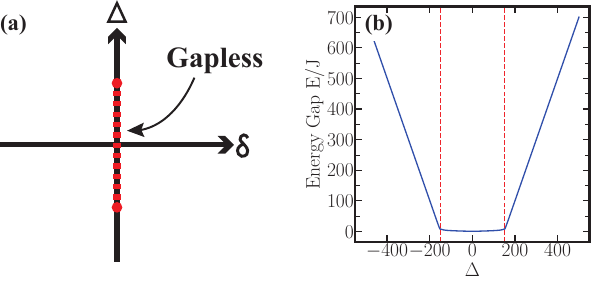}
    \caption{(a) Schematic representation of energy degeneracy locations using a mean-field approximation in parameter space. (b) The magnitude of the energy gap as a function of on-site potential.}
    \label{fig:mean_field}
\end{figure}

Building upon the original Hamiltonian expressed in momentum space, we consider the zeroth-order effective mean-field Hamiltonian, which describes many-particle states at a specific quasi-momentum \(k\). This mean-field Hamiltonian is constructed through a transformation applied to the basis of two different sublattices: \(\hat{a}_{k}^{\dag}/\hat{a}_{k} \rightarrow \sqrt{N_{a,k}}\) and \(\hat{b}_{k}^{\dag}/\hat{b}_{k} \rightarrow \sqrt{N_{b,k}}\). The zeroth order energy \(E_{0,k}\) is then given by:

\begin{equation}
 	\begin{split}
 	E_{0,k} =& 2JL\sqrt{n_{a}n_{0}-n_{a}^{2}}(1+\cos k) + 2\Delta_{a}n_{a}L +\\
 	 &\frac{UL}{2}(2n_{a}^{2}-2n_{a}n_{0}+n_{0}^{2}) +(-\Delta_{a}- \frac{U}{2})n_{0}L
 	\end{split}
\end{equation}

Here, \(n_{a} = \frac{N_{a}}{L}\) represents the density of the particles in the A sublattice in the mean-field limit, and \(n_{0} = \frac{N}{L}\) denotes the total particle density. Setting the condition \(\frac{\partial E_{k=0}}{\partial n_{a}} = 0\), we derive a quartic equation:

\begin{equation}
 	An_{a}^{4} + Bn_{a}^{3} + Cn_{a}^{2} + Dn_{a} + E = 0
  \label{quartic_eq}
\end{equation}

After performing meticulous calculations, the coefficients are found to be:

\begin{equation}
 	\begin{split}
 		A =& 4U^{2}L^{2} \\
 		B =& 4UL(2\Delta_{a}L-ULn_{0}) - 4U^{2}L^{2}n_{0} \\
 		C =& 4(J^{2}L^{2}(1+\cos k)^{2}) + (2\Delta_{a}L-ULn_{0})^{2} - \\
 		&4UL(2\Delta_{a}L-ULn_{0})n_{0}\\
 		D = & -4(J^{2}L^{2}(1+\cos k)^{2})n_{0}-(2\Delta_{a}L-ULn_{0})^{2}n_{0}\\
 		E = & (J^{2}L^{2}(1+\cos k)^{2})n_{0}^{2}
 	\end{split}
\end{equation}

Numerical solutions of this equation allow us to determine the value of \(n_{a}\), which is then used to construct the effective Hamiltonian within the zeroth-order mean-field approximation.

With the mean-field energy in hand, we construct the single-particle Hamiltonian. This is achieved by diagonalizing the Hamiltonian in each quasi-momentum \(k\) using the values \(n_{a}\) obtained from the numerical solution of \ref{quartic_eq}. As shown in Fig.~\ref{fig:mean_field}(b), the system exhibits a gapless line created by a series of degeneracy points, which are bounded by the values of $-n_{0}U/2$ and $n_{0}U/2$. This solution aligns with the results obtained from the effective Hamiltonian, where the boundaries are determined by the monopole that is farthest from the origin. 
Note that the stretching of a Dirac monopole, in a parameter space, from a point to a line has also been found in the mean-field treatment of an interacting two-component spinor Bose condensate system under the magnetic field~\cite{hoChernNumbersInteractionstretched2017}.

\section{Initial state is a fully occupying equally populated band insulating state }

Suppose we prepare the initial state of the system using the iTEBD method. In that case, it is critical to demonstrate that this state is a band-insulating state with a uniform weighting in the quasimomentum k space, which fulfills one of the prerequisites for quantized pumping. The initial state is prepared by evenly distributing $N$ particles in every unit cell. For $N$ particles in each unit cell, we consider the ground state obtained from iTEBD simulations at $t=0$ in the dimerized phase, with each unit cell being decoupled. This allows us to write down the full Hamiltonian $H$ as $\hat{H}_{\text{full}} = \bigotimes_{i=1}^{i=L} \hat{H}_{\text{local}}$ and the local Hamiltonian $\hat{H}_{\text{local}}$ is simply a two sites N-bosons problem which is explicitly written down, yielding a matrix with elements:

\begin{equation}
\begin{split}
H_{\text{ j,j}} = & \frac{U}{2}j(j-1) + \frac{U}{2}(n-j)(n-j-1) + \Delta j - \Delta(N-j), \\
H_{ \text{j, j+1/j-1}} = & J(\sqrt{N-j}\sqrt{j+1}).
\end{split}
\end{equation}

Solving the eigenstates $\psi_{\text{local}}$ of $H_{\text{local}}$ is equivalent to obtaining the whole spectrum of $H_{\text{full}}$. The entire wavefunction is denoted as $\Psi_{i} = 0\otimes0\otimes...\psi^{i}_{\text{local}}...\otimes0\otimes0$. In general, $\psi_{\text{local}}$ is simply a superposition state of the Fock basis of a single unit cell. Considering the particles are evenly filled in every unit cell in the dimerized phase, it will be the product state of all the same $\psi_{\text{local}}$ obtained by diagonalizing the two sites N-bosons Hamiltonian. As every unit cell reaches the same ground state through the iTEBD algorithm, the single-band state is guaranteed. The remaining task would be to prove the equal weighting of momentum $k$ in the $k$ space. 

For a unit cell with $M$ particles at site-A and $N-M$ particles at site-B, the general state is $\ket{\Psi} = (\hat{a}^{\dag}_{i})^{M}(\hat{b}^{\dag}_{i})^{N-M}\ket{0}$. A Fourier transform gives us the momentum representation of the initial state:

\begin{equation}
\ket{\Psi(k_{1},\ldots,k_{M},k'_{1},\ldots,k'_{N})} = \sum_{k_{1},\ldots,k_{M},k'_{1},\ldots,k'_{N}}\text{exp}\left(\mathbf{i} \bigg( \sum_{j=1}^{M}k_{j} + \sum_{j=1}^{N-M}k'_{j} \bigg) i \right)(\hat{a}^{\dag}_{k_{1}}\ldots\hat{a}^{\dag}_{k_{M}})(\hat{b}^{\dag}_{k'_{1}}\ldots\hat{b}^{\dag}_{k'_{N-M}})\ket{0}.
\label{multi_k_state}
\end{equation}

The phase factors for individual particles in the Brillouin zone form a $\mathbb{Z}_{L}$ cyclic group, defined as $\mathbb{Z}_{L} = \{ e^{\mathbf{i}ki}, k = \frac{2\pi}{L}j \wedge j \in \mathbb{Z}: j \in [0, L-1] \}$, with $L$ being the number of unit cells. The direct product of such groups in the 2-particle case forms the phase factors in Eq.~(\ref{multi_k_state}), given by $\mathbb{Z}_{L}\times\mathbb{Z}_{L} = \{ e^{\mathbf{i}(k_{1}+k_{2})i}, {j,s} \in [0, L-1] \}$. Fixing $k_{1}$ and summing over all possible values of $k_{2}$, the product is well-defined in the Brillouin zone with a $2\pi$ phase shift. This summing approach can be extended to multi-particle states, provided the state is a single Fock state. Thus, a multi-particle momentum vector $k$ is defined by summing all single-particle momenta $k_{i}$ with equal probability within the multi-particle Brillouin zone.

If the initial state is a superposition of Fock bases in the same unit cell, the equal weight property in k is guaranteed.  The expectation is that it should embody a uniformly occupied band insulating state, which is crucial for the quantized pumping process.

\section{Pumping obstacle induced by the multi-band degeneracy in SU(2) model  }

\begin{figure}[h]
    \centering
    \includegraphics[scale=0.9]{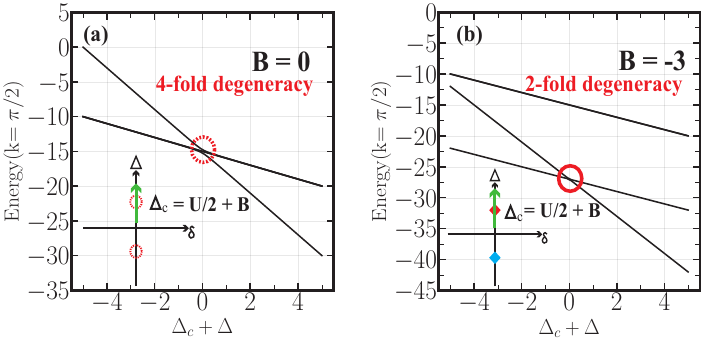}
    \caption{The lowest few energy spectra along the $\Delta$ axis in the parameter space around the degeneracy point. The system setting is U = 30, J = 0.1 (a) without the staggered Zemann-field (B=0) (b) with the staggered Zemann-field splitting (B=-3) }
    \label{SU2_monopole}
\end{figure}

As mentioned in the main text, there is a main obstacle in preparing a single non-degenerate energy band during the entire adiabatic evolution. In the fermionic system, the degeneracy comes from the permutation of spin species in real space. This is illustrated in Fig.~\ref{SU2_monopole}, where we have performed calculations on a 2-cell system with $U = 30$ and $J = 0.1$. As depicted in Fig.~\ref{SU2_monopole}(a) and (b), in the limit where $J \ll U$ (with $J = 0.1$), three energy bands are closely degenerate, even though the band with uniform occupation is well prepared. This closeness in energy also poses a challenge to the adiabatic pumping process. However, introducing a staggered field achieves the two lowest non-degenerate bands.

\section{Effective Hamiltonian of an SU(3) model}
We construct an unperturbed Hamiltonian in real space by selecting an appropriate basis. When the Zeeman field is included, the two lowest nearly degenerate energy bands split into two non-degenerate bands. The chosen basis set is $\{ \hat{c}^{\dag}_{i,\alpha} \ket{\text{G.S.}}, i \in [1,L] \}$, where $\hat{c}^{\dag}_{i,\alpha}$ is the creation operator for a specific type of fermion-$\alpha$ at site $i$, and $\ket{\text{G.S.}}$ denotes the ground state with two other types of fermions filling up the lattice to achieve the lowest energy state.

Considering the Zeeman coupling term $H_{\text{Zeeman}} = B\sum_{i}\hat{S}^{z}_{s=1}$, the unperturbed Hamiltonian can be readily constructed. Assuming we have three types of fermions denoted as $\{ \alpha, \beta, \gamma \}$, the corresponding coupling strengths are $(B,0,-B)$. Focusing on fermion-$\alpha$, which has the highest Zeeman coupling strength $B$, hopping in the background of a two-particle filling per unit cell filled up by the other two types of fermion species, the lowest energy states of the background $\ket{\text{G.S}}$ can be easily determined. It would be the state where fermion-$\beta$ and fermion-$\gamma$ fill up the lower potential sites with the Zeeman field pointing in the positive direction. Consequently, by applying perturbation theory, the real-space Hamiltonian can be explicitly written as:

\begin{equation}
H = \begin{pmatrix}
    E_{a} & J_{1} & 0 & \dots & 0 \\
    J_{1} & E_{b} & J_{2} & \dots & 0 \\
    0 & J_{2} & E_{a} & \ddots & \vdots \\
    \vdots & & \ddots & \ddots & J_{L-1} \\
    0 & & \dots & J_{L-1} & E_{b}
\end{pmatrix},
\end{equation}
where $E_{a} = L(-2\Delta_{0} - B + U) + \Delta_{0} - B$, and $E_{b} = (L - 1)(-2\Delta_{0} - B + U) + (-3\Delta_{0} + 3U)$.

Applying the Fourier transform to transition to momentum space, the Hamiltonian is reconstructed in block form:
\begin{equation}
H(k) = \begin{pmatrix}
        E_{a} & J_{1} + J_{2}e^{\mathbf{i}k} \\
        J_{1} + J_{2}e^{-\mathbf{i}k} & E_{b}
    \end{pmatrix}.
\end{equation}

Upon solving for the energy spectrum, we observe that $\Delta = U + B$ is the sole degeneracy point contributing to the constructed Hamiltonian. Using the same methodology, other monopoles are obtained in a similar way.

%\bibliography{SM}% Produces the bibliography via BibTeX.
%
% ****** End of file apssamp.tex ******

\end{document}